\documentclass[journal,comsoc]{IEEEtran}

\usepackage[T1]{fontenc}
\usepackage{xspace,amsmath,amssymb,amsfonts,epsfig,syntonly,bbm,dsfont}
\usepackage{cite,bm,color,url,textcomp,empheq,boxedminipage}
\usepackage{algorithmicx,algorithm,algpseudocode}
\usepackage{epstopdf}
\usepackage{empheq}
\usepackage{graphicx,graphics,subfigure}
\usepackage{multirow,multicol}
\usepackage{psfrag}
\usepackage{stfloats}

\newtheorem{theorem}{{{Theorem}}}
\newtheorem{lemma}{{Lemma}}
\newtheorem{proposition}{{Proposition}}
\newtheorem{remark}{{Remark}}

\DeclareMathOperator*{\argmin}{arg\,min}

\long\def\symbolfootnote[#1]#2{\begingroup
\def\thefootnote{\fnsymbol{footnote}}
\footnote[#1]{#2}\endgroup}

\psfull

\begin{document}
\title{Sequential Offloading for Distributed DNN Computation in Multiuser MEC Systems}

\author{Feng Wang,~\IEEEmembership{Member, IEEE}, Songfu Cai,~\IEEEmembership{Member, IEEE}, and Vincent K. N. Lau,~\IEEEmembership{Fellow, IEEE}

\thanks{F. Wang is with the School of Information Engineering, Guangdong University of Technology, Guangzhou 510006, China, and is also with the Department of Electronic and Computer Engineering, The Hong Kong University of Science and Technology, Hong Kong (e-mail: fengwang13@gdut.edu.cn).}

\thanks{S. Cai and V. K. N. Lau are with the Department of Electronic and Computer Engineering, The Hong Kong University of Science and Technology, Hong Kong (e-mail: scaiae@connect.ust.hk, eeknlau@ust.hk).}
\vspace{-0.4cm}
}

\maketitle

\begin{abstract}
  This paper studies a {\em sequential} task offloading problem for a multiuser mobile edge computing (MEC) system. While most of the existing works consider static one-shot offloading optimization with fixed wireless channel conditions and fixed computational tasks, we consider a dynamic optimization approach, which embraces wireless channel fluctuations and random deep neural network (DNN) task arrivals over an infinite horizon. Specifically, we introduce a local CPU workload queue (WD-QSI) and an MEC server workload queue (MEC-QSI) to model the dynamic workload of DNN tasks at each WD and the MEC server, respectively. The transmit power and the partitioning of the local DNN task at each WD are dynamically determined based on the instantaneous channel conditions (to capture the transmission opportunities) and the instantaneous WD-QSI and MEC-QSI (to capture the dynamic urgency of the tasks) to minimize the average latency of the DNN tasks. The joint optimization can be formulated as an ergodic Markov decision process (MDP), in which the optimality condition is characterized by a centralized Bellman equation. However, the brute force solution of the MDP is not viable due to the curse of dimensionality as well as the requirement for knowledge of the global state information. To overcome these issues, we first decompose the MDP into multiple lower dimensional sub-MDPs, each of which can be associated with a WD or the MEC server. Next, we further develop a parametric online Q-learning algorithm, so that each sub-MDP is solved locally at its associated WD or the MEC server. The proposed solution is completely decentralized in the sense that the transmit power for sequential offloading and the DNN task partitioning can be determined based on the local channel state information (CSI) and the local WD-QSI at the WD only. Additionally, no prior knowledge of the distribution of the DNN task arrivals or the channel statistics will be needed for the MEC server. The proposed solution can achieve the superb performance over various state-of-the-art baselines.
 \end{abstract}


\section{Introduction}
 Modern applications of Internet-of-things (IoT), such as augmented reality/virtual reality (AR/VR), demand real-time inferencing of deep neural networks (DNNs). The DNN can effectively extract high-level features from the raw data at high computational complexity cost\cite{Chiang16,DNN19,DL-theory21}. The input data for the DNN in these new applications are generated locally at the mobile devices, and hence it is natural for the DNN inferencing tasks to be computed locally. However, it is very challenging to compute the DNN tasks at the mobile devices due to the limited computation and battery capacity\cite{ML-Facebook19}. Therefore, it is important to exploit external computation resources to realize the full potential and benefits of future device-based artificial intelligence (AI) applications.

 Mobile edge computing (MEC) has been proposed to provide cloud-like computational offloading services in the proximity of data sources\cite{Mach17,JunZhang17,Feng18,Feng19,Taleb17}. In MEC systems, resource-constrained wireless devices (WDs) can offload all or part of their computation tasks to the network edge nodes, such as cellular base stations (BSs) or access points (APs), and then the integrated MEC servers therein can execute these offloaded tasks on behalf of the WDs. Inspired by MEC systems, the computational intensive DNN task can be partitioned into a {\em front-end task} for local computation at the WDs and a {\em back-end task} for offloading to the MEC server. Leveraging the advances in MEC technologies\cite{EdgeInte-ProcIEEE19}, the real-time computation of complex DNNs on WDs becomes feasible.

 In the literature, some recent works exploit MEC for computational offloading of DNN tasks under a single WD setup \cite{Chen21,Neurosurgeon17,ACM18,DNN-MEC-GC20,
 ICDSC20,EdgeAI-TWC20,CoEdge-TIN21,ICDCS17,Infocom19,Infocom20,Auto-2021} or a multi-WD setup\cite{CoEdge-TIN21,Multiuser-IOTJ21}. The DNN task offloading and resource allocation schemes are designed to optimize the WD's energy consumption\cite{DNN-MEC-GC20,Infocom19,ICDCS17}, the DNN inferencing accuracy\cite{EdgeAI-TWC20,ICDSC20}, and the DNN inferencing time\cite{Auto-2021,ACM18,Infocom20,Chen21}. The profiling knowledge of layer-wise DNN inferencing delay/energy consumption, which heavily depends on the MEC system parameters, is determined in either an offline manner \cite{DNN-MEC-GC20,Infocom19,ICDCS17,ACM18,Infocom20,ICDSC20,EdgeAI-TWC20} or by an online learning approach\cite{Auto-2021,Chen21}. The computational offloading schemes for DNN inferencing tasks among multiple WDs have been considered to minimize the system energy consumption \cite{CoEdge-TIN21} or the computation delay\cite{Multiuser-IOTJ21}. However, in all these works \cite{Auto-2021,ACM18,Infocom20,DNN-MEC-GC20,
 ICDSC20,Infocom19,ICDCS17,EdgeAI-TWC20,Multiuser-IOTJ21,CoEdge-TIN21,Chen21}, the MEC offloading for DNN inferencing tasks is all based on one-shot optimization, where the DNN inferencing tasks and channel conditions are assumed to be static. In practice, the DNN inferencing tasks arrive sequentially in a random manner\cite{DNN19,DL-theory21,ML-Facebook19}. Furthermore, a DNN task may have to span multiple time slots to finish. In addition, the wireless channels are also time-varying due to the mobility of the WDs\cite{Chai-19,Goldsmith-Book}. Thus, the existing static optimization approach fails to embrace the dynamic nature of the DNN task offloading problem.

 In this paper, we consider a multiuser MEC system, where the DNN task randomly arrives at the WDs over an infinite horizon. Each DNN task may span across multiple slots. We consider a dynamic optimization formulation, which embraces wireless channel fluctuations and random DNN task arrivals over an infinite horizon to minimize the long term average computational latency of the DNN tasks, subject to the energy constraints. The key contributions of this paper are summarized as follows.

 \begin{itemize}
 \item{\bf Sequential Offloading Design under an MDP Framework:} We introduce a local CPU workload queue (WD-QSI) and an MEC server workload queue (MEC-QSI) to model the dynamic workload at each WD and the MEC server, respectively. The transmit power for sequential offloading and the partitioning of the local DNN task at each WD are dynamically determined based on the instantaneous channel conditions (to capture the transmission opportunities) and the instantaneous WD-QSI and MEC-QSI (to capture the dynamic urgency of the tasks) to minimize the average energy consumption and computation latency of the DNN tasks. The joint optimization of the local task partitioning, the transmit power control for offloading, and the local/remote computation is formulated as an ergodic Markov decision process (MDP), in which the optimality condition is characterized by a centralized Bellman equation.

 \item{\bf Decentralized MDP Decomposition:} The brute force solution of the MDP \cite{Goodfellow-2016,FL-off20,FedAdapt,Split-Learning18,Privacy20,Sutton2018} becomes not viable due to the curse of dimensionality (i.e., a large number of WDs), as well as the requirement for knowledge of the global channel state information (CSI) and the global WD-QSI/MEC-QSI. To overcome these challenges, we propose a novel decentralized MDP decomposition approach. Specifically, by exploiting the problem-specific structure (i.e., the additive structure of the cost function and the independency among the WDs in the transmit power constraints for task offloading and local computing), the centralized Bellman optimality equation is decomposed into multiple lower dimensional sub-MDPs, each of which can be associated with a WD or the MEC server. The proposed solution is completely decentralized in the sense that the transmit power and the DNN task partitioning can be determined based on the local CSI and the local WD-QSI at the WD only.

 \item {\bf Parametric Online Q-Learning Solution:} Note that the CSI and WD-QSI/MEC-QSI are switching in continuous state spaces with uncountably many realizations. As a result, the standard Q-learning methods \cite{Sutton2018,Goodfellow-2016}, which utilize a lookup table to store the Q-values for the finite state-action pairs, cannot be adopted to solve the decentralized Bellman optimality equation of each sub-MDP. As such, we further propose a new parametric online Q-learning algorithm, so that each sub-MDP can be solved locally at its associated WD or the MEC server. The Q-function associated with the sub-MDP is parameterized by a weighted sum of sigmoid functions. Based on the gradient descent (GD) method, the weights of each Q-function are trained online by minimizing the residual error of the Bellman equation for the corresponding sub-MDP. We provide the guaranteed convergence analysis of the proposed solution. Additionally, no prior knowledge of the distribution of the DNN task arrivals or the channel statistics will be needed for the MEC server. The proposed solution can achieve the superb performance over various state-of-the-art baselines.

 \end{itemize}

 The remainder of the paper is organized as follows. Section II introduces the multiuser MEC system model with sequential DNN task offloading. Section III formulates the infinite horizon discounted MDP problem to minimize the long-term expected discounted cost, by jointly optimizing multiuser DNN task partitioning, computation offloading, and local/remote computing. Section IV presents the decentralized MDP decomposition. Section V develops the parametric online Q-learning algorithm. Section VI provides numerical results to demonstrate the effectiveness of the proposed design, which is followed by the conclusion in Section VII.


\section{System Model}\label{Sec:System}

\begin{figure}
  \centering
  \includegraphics[width = 3.5in]{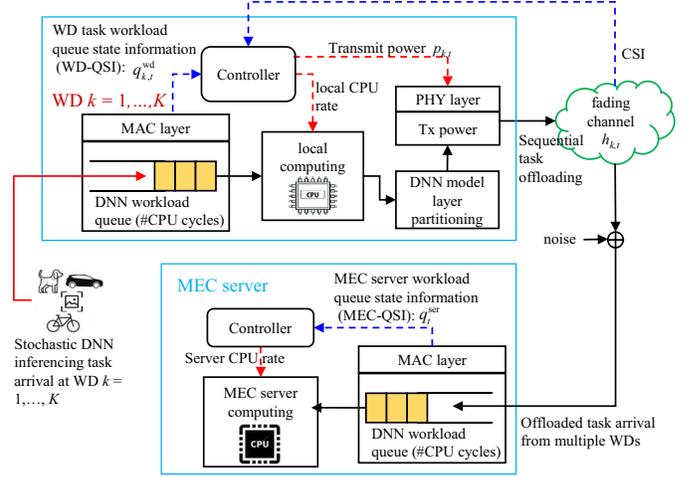}
 \caption{System model of a multiuser MEC system with the stochastic DNN inferencing task arrivals and sequential offloading.} \label{fig.DNN-model-fig}
\end{figure}

\begin{figure}
  \centering
  \includegraphics[width = 3.5in]{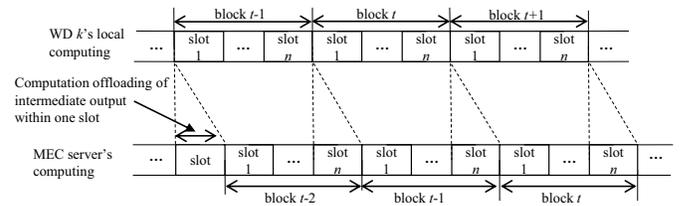}
 \caption{Proposed timeline for the local computing of the WDs and the remote computing of the MEC server.} \label{fig.timeline}
\end{figure}

 We consider a multiuser MEC system, which consists of $K$ WDs and a BS integrated with an MEC server. As illustrated in Fig.~\ref{fig.DNN-model-fig}, each WD $k\in{\cal K}\triangleq\{1,...,K\}$ acts as a controller, which takes the local CPU workload of the DNN task arrivals (i.e., WD-QSI) and the CSI for sequential offloading as input, and generates the local CPU rate (i.e., the frequency of running CPU cycles) of the local computing and the transmits power of offloading as output. The MEC server controls its CPU rate to execute the offloaded DNN inferencing tasks from the $K$ WDs. As shown in Fig.~\ref{fig.timeline}, the multiuser MEC system operates in a slotted structure, where the time dimension is partitioned into a decision block indexed by $t\in{\cal T}\triangleq \{0,1,...\}$. Each decision block consists of $n\geq 1$ slots and each slot duration is $\tau$.

 \subsection{DNN Task Workload Queuing at the WDs}

 We assume that a DNN inferencing task arrives at each WD $k\in{\cal K}$ at the beginning of each slot. Let the binary variable $\beta_{k,t,i}\in\{0,1\}$ denote whether one DNN task arrives or not at the $i$th slot of block $t$, where $\beta_{k,t,i}=1$ represents that one DNN inferencing task arrives at the WD. The random binary sequence $\{\beta_{k,t,i}\}$ is independent and identically distributed (i.i.d.) over the slots. We further assume that $\beta_{k,t,i}$ follows a Bernoulli distribution with probability $b_k\in(0,1)$, i.e., ${\rm Pr}(\beta_{k,t,i}=1)=b_k$ and ${\rm Pr}(\beta_{k,t,i}=0)=1-b_k$ for $i=1,...,n$ and $k\in{\cal K}$. Each DNN inferencing task is assumed to be successfully executed after running a number of $W$ CPU cycles by local/remote computing. Therefore, we treat each DNN inferencing task as a fixed workload of $W$ CPU cycles.

 As shown in Fig.~\ref{fig.DNN-model-fig}, each WD $k\in{\cal K}$ maintains a local first-in-first-out (FIFO) workload queue for the arrived but not yet executed DNN inferencing tasks. Denote by $q^{\text{wd}}_{k,t}$ the WD-QSI (in the unit of CPU cycles) at the CPU workload queue of WD $k$ at the beginning of the $t$th block. Then, the CPU workload queue dynamics of local DNN inferencing computation at each WD $k\in{\cal K}$ evolves according to
 \begin{align}\label{eq.Q-wdk}
    q^{\text{wd}}_{k,t+1} = \Big[q^{\text{wd}}_{k,t} - \sum_{i=1}^n f_{k,t,i}^{\text{wd}}\tau\Big]^+ +\sum_{i=1}^n\beta_{k,t,i}W,
 \end{align}
 where $[x]^+\triangleq \max(x,0)$, $t\in{\cal T}$ denotes the block index, $f_{k,t,i}^{\text{wd}}$ denotes the CPU rate of WD $k$'s local computing during the $i$th slot of block $t$, and the term $\sum_{i=1}^n\beta_{k,t,i}W$ denotes the new arrival workload of DNN inferencing tasks within the $n$ slots of the $t$th block.

 Within each block $t\in{\cal T}$, each WD $k\in{\cal K}$ performs the local computing to execute the front-end part of the DNN tasks. As the CPU chip voltage is approximately linear to the CPU rate within the CPU operational voltage range, the power consumption of the WD $k$'s CPU is thus proportional to the third power of the CPU rate, i.e., $(f^{\text{wd}}_{k,t,i})^3$\cite{Mach17,JunZhang17}. Accordingly, the amount of energy consumption due to the local computing of WD $k\in{\cal K}$ at the $t$th block is expressed as \cite{JunZhang17}
 \begin{align}\label{eq.ek-loc}
    E_{k,t}^{\text{wd}} = \sum_{i=1}^n\tau \xi^{\text{wd}}_k\left(f^{\text{wd}}_{k,t,i}\right)^3,
 \end{align}
 where $\xi^{\text{wd}}_k>0$ denotes the effective switched capacitance of the CPU architecture of WD $k$.

\subsection{Computation Offloading from the WDs to the MEC Server}
 At the end of the $t$th decision block, each WD $k\in{\cal K}$ generates the intermediate output of one DNN inferencing task after its local computing. In order for the MEC server to execute the remaining workload of one DNN inferencing task after the WD $k$'s local computing per block, WD $k$ needs to transmit the respective intermediate output (i.e., the computed results of the front-end part of one DNN inferencing task) to the MEC server via the computation offloading. Therefore, the MEC server is responsible for computing the remaining back-end part of one DNN inferencing task. As shown in Fig.~\ref{fig.timeline}, the computation offloading procedure of the intermediate output is considered to be completed within one slot.

 In order to avoid co-channel interference, we consider the frequency-division multiple access (FDMA)-based protocol for the multiuser computation offloading from the $K$ WDs to the BS, where each WD $k\in{\cal K}$ is allocated a dedicated resource block of $B_k$ spectrum bandwidth (in Hz) at the slot just after the $t$th local-computing block of WD $k$. Let $h_{k,t}>0$ and $p_{k,t}>0$ denote the wireless channel power gain and the transmit power, respectively, for WD $k$'s computation offloading at the slot just after the $t$th local-computing block. By assuming that the additive white Gaussian noise (AWGN) at the BS receiver is with unit variance, the data rate (bits/s) for task offloading from WD $k\in{\cal K}$ to the BS at the slot just after the $t$th local-computing block is written as
 \begin{align}\label{eq.rk}
    r_{k,t} = B_k\log_2\left(1+\frac{h_{k,t}p_{k,t}}{\Gamma}\right),
 \end{align}
 where $\Gamma\geq 1$ denotes the gap of the signal-to-noise ratio (SNR) due to the employed modulation and coding scheme (MCS). Regarding the channel power gain $h_{k,t}$, we consider that $h_{k,t}$ remains constant within the slot between the $t$th block of local computing and the $t$th block of the MEC server's computing, and it follows an i.i.d. exponential distribution with the large-scale path loss as the mean value\cite{Goldsmith-Book}.

 Denote by $D_{k,t}$ the intermediate output size (in bits) of the local computing of WD $k$ at the $t$th block. It is assumed that the intermediate output size $D_{k,t}$ of the local computing is proportional to the remaining DNN inferencing computation workload to be executed. In this case, the intermediate output size $D_{k,t}$ can be expressed as a function of the local-computing CPU rate vector $\bm f_{k,t}^{\text{wd}}$ of WD $k$ at the $t$th block, i.e., $D_{k,t}(\bm f_{k,t}^{\text{wd}}) = \zeta\Big(\big[q^{\text{wd}}_{k,t} - \sum_{i=1}^n f_{k,t,i}^{\text{wd}}\tau\big]^+~{\rm mod}~W\Big)$, where $\zeta>0$ is a positive constant factor, and $\bm f_{k,t}^{\text{wd}}\triangleq [f_{k,t,1}^{\text{wd}},...,f_{k,t,1}^{\text{wd}}]^{\tt T}$, with the superscript ${\tt T}$ denoting the transpose operation. Note that the intermediate output of one DNN task not yet completely executed should be successfully offloaded to the MEC server within one slot. As a result, it follows that $r_{k,t}\tau\geq D_{k,t}(\bm f_{k,t}^{\text{wd}})$. Based on \eqref{eq.rk}, the transmit power for computation offloading is expressed as
 \begin{align}\label{eq.tx-power}
   p_{k.t} =
   \frac{\Gamma}{h_{k,t}}\Big(2^{\frac{D_{k,t}(\bm f_{k,t}^{\text{wd}})}{\tau B_k}}-1 \Big),
 \end{align}
 where $k\in{\cal K}$. Denoting by $E_{k,t}^{\text{off}}$ the amount of energy consumption of WD $k$ incurred by offloading the intermediate output to the BS within the slot after the $t$th block, we have
 \begin{align}\label{eq.ek-off}
   E_{k,t}^{\text{off}} = \tau p_{k,t},~\forall k\in{\cal K}.
 \end{align}

\subsection{DNN Task Workload Queuing at the MEC Server}
 After the BS has received the intermediate output of the partitioned DNN inferencing task from the WDs, the MEC server (integrated with the BS) starts to process the back-end DNN tasks. The MEC sever maintains a FIFO workload queue for the back-end DNN tasks. Let $q^{\text{ser}}_t$ denote the MEC-QSI (in the unit of CPU cycles) at the CPU workload queue of the MEC server at the beginning of the $t$th block. As shown in Fig.~\ref{fig.timeline}, the timeline of the MEC server is one-block slower than that of the WDs, i.e., the $t$th block at the MEC server is delayed for $(n+1)$ slots when compared to the $t$th block at the WDs. Let $f^{\text{ser}}_{t,i}$ denote the CPU-cycle frequency assigned to execute the back-end parts of the WDs' DNN inferencing tasks at the $i$th slot of the $t$th remote-computing block, where $i=1,...,n$. The DNN inferencing computation workload queue dynamics at the MEC server is given by
 \begin{align}\label{eq.Q-server}
 q_{t+1}^{\text{ser}} &= [q_{t}^{\text{ser}}-\sum_{i=1}^nf^{\text{ser}}_{t,i}\tau]^+ \notag \\
 &~~ + \sum_{k=1}^K \Big(\big[q^{\text{wd}}_{k,t} - \sum_{i=1}^n f_{k,t,i}^{\text{wd}}\tau\big]^+~{\rm mod}~W\Big),
 \end{align}
 where $(\big[q^{\text{wd}}_{k,t} - \sum_{i=1}^n f_{k,t,i}^{\text{wd}}\tau\big]^+{\rm mod}~W)$ represents the amount of new DNN inferencing computation workload arrivals at the MEC server at the end of the $t$th remote-computing block, with $a~{\rm mod}~b$ being a modulo operator. Accordingly, the amount of energy consumption incurred by the MEC server's execution at the $t$th remote-computing block is given by
 \begin{align}
    E^{\text{ser}}_t = \sum_{i=1}^n \tau \xi^{\text{ser}} (f^{\text{ser}}_{t,i})^3,
 \end{align}
 where $\xi^{\text{ser}}>0$ denotes the effective switched capacitance of the CPU architecture at the MEC server.

\section{Problem Formulation}
 In this section, by taking into account the stochastic DNN inferencing task arrivals and time-varying channel fading for offloading, we formulate the optimal dynamic DNN inferencing task partition and resource allocation as an infinite horizon discounted MDP. We also characterize the optimize solution using the Bellman optimality equation.

\subsection{System State and Control Action}
 Denote by $s_t=(q^{\text{ser}}_t,\{q_{k,t}^{\text{wd}},\forall k\},\{h_{k,t},\forall k\},\{\beta_{k,t,i},i=1,...,n,\forall k\})\in{\cal S}$ the overall system state at the $t$-th block under consideration, where $\cal S$ is the system state space. Denote by ${\cal A}$ the action space of the multiuser MEC system. Let $\pi:{\cal S}\mapsto{\cal A}$ denote the control policy that makes control decisions based on the state $s_t$. We use $a_{\pi}(s_t) = (\{p_{k,t},\forall k\}, \{\bm f_{k,t}^{\text{wd}},\forall k\in{\cal K}\}, \bm f^{\text{ser}}_{t})\in{\cal A}$ to denote the control action taken when the system is in state $s_t$ under a stationary control policy $\pi$, where $\bm f_{k,t}^{\text{wd}} \triangleq [f^{\text{wd}}_{k,t,1},...,f^{\text{wd}}_{k,t,n}]^{\tt T}$ and $\bm f^{\text{ser}}_{t}\triangleq [f^{\text{ser}}_{t,1},...,f^{\text{ser}}_{t,n}]^{\tt T}$. This control action $a_\pi(s_t)$ causes the system to transition to a new state $s_{t+1}$. It is worth noting that both the state space and action space are continuous.

\subsection{Infinite-Horizon MDP Problem Formulation}
 In this paper, we are interested in designing a control policy $\pi$ to minimize the long-term expected discounted sum cost from any initial state $s\in{\cal S}$. Let
 \begin{align}\label{eq.c_a_pi}
 c_{a_{\pi}}(s_t) \triangleq  w_1 q_{t}^{\text{ser}} + w_2 E_t^{\text{ser}} + \sum_{k=1}^K \Big(w_3 q_{k,t}^{\text{wd}} + w_4 E_{k,t}\Big)
 \end{align}
 denote the weighted sum of DNN inferencing computation workload queues and computation energy cost incurred by a control action $a_{\pi}(s_t)$ at the $t$th block, where $E_{k,t}\triangleq E^{\text{wd}}_{k,t}+E^{\rm off}_{k,t}$ is defined as the WD $k$'s total energy consumption of local computing at the $t$th block (c.f.~\eqref{eq.ek-loc}) and the subsequent computation offloading in one slot (c.f.~\eqref{eq.ek-off}), and $\{w_1,w_2,w_3,w_4\}$ denote the non-negative weights specified for different system design preferences. Starting from any initial state $s\in{\cal S}$, we consider the long-term expected discounted sum cost minimization problem as
 \begin{subequations}\label{eq.v-pi}
 \begin{align}
 ({\rm P1}):~&\min_{\pi} J^{\pi}(s) \triangleq \lim_{T\rightarrow \infty}\mathbb{E}^{\pi}\Big\{ \sum_{t=0}^T\gamma^t c_{a_{\pi}}(s_t)\Big| s_0=s\Big\} \\
 &~~{\rm s. t.}~~p_{k.t} =
   \frac{\Gamma}{h_{k,t}}\Big(2^{\frac{D_{k,t}(\bm f_{k,t}^{\text{wd}})}{\tau B_k}}-1 \Big),~\forall k, \forall t  \\
 &~~~~~~~~0 \leq f^{\text{wd}}_{k,t,i}\leq f^{\text{wd}}_{k,\max},~i=1,...,n, \forall k,\forall t \\
 &~~~~~~~~0 \leq f^{\text{ser}}_{t,i}\leq f^{\text{ser}}_{\max},~i=1,...,n,\forall t,
 \end{align}
 \end{subequations}
 where $0<\gamma<1$ is the discount factor to determine how much the costs in the distant future are relative to those in the immediate future. The expectation in (\ref{eq.v-pi}a) is taken over different decision makings under different system states following a control policy $\pi$ across the decision blocks. The objective function $J^{\pi}(s)$ in problem (P1) is also referred to as the state value function of the system state $s_0$ under a control policy $\pi$\cite{Sutton2018}.

 The random state process $\{s_t\}$ is a controlled Markov process with the transition kernel given by
 \begin{align} \label{eq.central-tx-kernel}
  &{\rm Pr}(s_{t+1}|s_t,a_{\pi}(s_t)) = \Big( \prod_{k=1}^K {\rm Pr}(h_{k,t+1})\prod_{i=1}^n{\rm Pr}(\beta_{k,t+1,i})\Big) \notag \\
  &~~~~~~~~~~~~~~~~~~ \times {\rm Pr}(q^{\text{ser}}_{t+1},\{q_{k,t+1}^{\text{wd}},\forall k\} | s_t,a_{\pi}(s_t)).
 \end{align}
 Based on \eqref{eq.central-tx-kernel}, problem (P1) is an infinite horizon discounted MDP problem. Define $Q^{\pi}(s,a)$ as the expected value of taking action $a$ in state $s$ and thereafter following the policy $\pi$, which is expressed as
 \begin{align}
 Q^{\pi}(s,a) = \lim_{T\rightarrow \infty}\mathbb{E}^{\pi}\left\{
 \sum_{t=0}^T\gamma^t c_{a_{\pi}}(s_t) \Big| s_0=s,a_0=a
 \right\}.
 \end{align}
 Note that $Q^{\pi}(s,a)$ is referred to as the action-value function or the Q-function. Denote by $\pi^*$ the optimal solution of problem (P1), which is characterized by the Bellman optimality equation\cite{Sutton2018}. This is summarized in the following lemma.

 \begin{lemma}[Optimality Conditions of MDP~(P1)]\label{lemma.P1-Bellman}
 At the optimality of the MDP problem (P1), we have the Bellman optimality equation for the action-value function $Q^{\pi}(s,a)$:
 \begin{align}\label{eq.Bellman-eq}
 Q^*(s,a) &= \mathbb{E}\Big\{c(s,a) + \gamma \min_{a^\prime} Q^{*}(s_{t+1},a^\prime) \Big| s_t=s,a_t=a\Big\},
 \end{align}
 where $c(s,a) \triangleq c_{a_{\pi}}(s)$ and $Q^*(s,a)\triangleq Q^{\pi^*}(s,a)$ are defined for notation conciseness, $\pi^* = \argmin_{\pi}~Q^{\pi}(s,a)$ denotes the optimal policy, and the expectation in \eqref{eq.Bellman-eq} is taken over the state transition probabilities of $\{s_t\}$ (i.e., ${\rm Pr}(s_{t+1}|s_t,a_t=a_{\pi^*}(s_t))$, $\forall s_{t+1}\in{\cal S}$). 
 \end{lemma}

\begin{remark}
 For the Bellman optimality equation \eqref{eq.Bellman-eq} in Lemma~\ref{lemma.P1-Bellman}, it is quite challenging to obtain the optimal control action $\pi^*$ based on the dynamic programming (DP) methods. The reasons are stated as follows. First, due to the {\em curse of dimensionality} (i.e., a large dimension of system state space ${\cal S}$, which involves a total of $(K+1)$ workload queues, $K$ channel conditions, and $Kn$ task arrivals), it is usually of high time-complexity to compute the Bellman optimality equation \eqref{eq.Bellman-eq}. This is not appropriate for timely decision makings in the time-sensitive DNN interfering applications. Second, solving \eqref{eq.Bellman-eq} involves a significant signaling overhead by collecting the global CSI $\{h_{k,t}\}$ and the global WD-QSI $\{{q}^{\text{wd}}_{k,t}\}$ of the $K$ WDs at each decisional time block $t\in{\cal T}$. Third, the system state space $\cal S$ and control action $a_{\pi}(s_t)$ are both continuous, which causes \eqref{eq.Bellman-eq} not to be numerically solved by the value iteration or policy iteration based reinforcement learning methods without introducing state/action discretization.
\end{remark}

 From an implementation perspective, it is desirable to obtain a decentralized solution, in which each WD performs its decision makings only based on the local WD-QSI and the local CSI. To address the above challenges, we next pursue to obtain a decentralized solution for (P1) based on the equivalent decomposition and parametric online Q-learning approach.

\section{Decentralized Decomposition for (P1)}
\begin{figure}
  \centering
  \includegraphics[width = 3.5in]{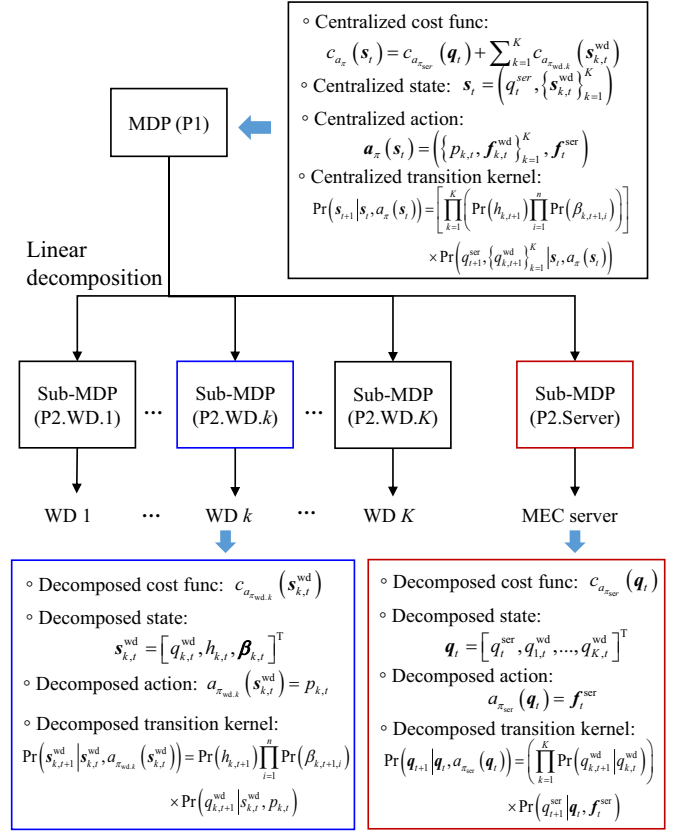}
 \caption{The decomposition of the MDP (P1).} \label{fig.Decom}
\end{figure}

 In this section, we exploit the problem-specific structure to decompose the centralized MDP (P1) into a total of $(K+1)$ sub-MDPs, which respectively correspond to the $K$ WDs and the MEC server, as illustrated in Fig.~\ref{fig.Decom}. Specifically, via exploiting the linear additive structure, we can decompose the objective function of (P1) into a total of $(K+1)$ parts, each of which is the cost function of one WD or the MEC server. Furthermore, leveraging the independency among the $K$ WDs' transmit power for task offloading and local computing, the constraints of (P1) are decomposed into the individual actions of the WDs and the MEC server. With such a decentralized decomposition for (P1), each WD can individually make its decision based on the local WD-QSI and local CSI, and the MEC server will not require the prior knowledge of the random task arrivals or the global CSI.


\subsection{Decomposition of Problem (P1)}
 Note that, across all the decision blocks $t\in{\cal T}$, each WD $k\in{\cal K}$ firstly makes its decision based on the local WD-QSI and the local CSI, and then the MEC server makes its decision based on both the MEC-QSI at the MEC server and the global WD-QSI of the $K$ WDs. As such, the decomposition of problem (P1) can dramatically reduce the total number of decision makings and simplify the signaling overhead in the decentralized implementation procedure.

 First, we introduce the definitions as follows. For the MEC server, it follows that
 \begin{itemize}
 \item MEC server state space: ${\cal Q} =\{\bm q_{0},\bm q_{1},...\}$, where $\bm q_{t} \triangleq [q^{\text{ser}}_{t},q^{\text{wd}}_{1,t},...,q^{\text{wd}}_{K,t}]\in\mathbb{R}^{(K+1)\times 1}$;
 \item MEC server action space: ${\cal A}^{\text{ser}}=\{\bm f^{\text{ser}}_{0},\bm f^{\text{ser}}_{1},...\}$, where $\bm f^{\text{ser}}_{t}= [f^{\text{ser}}_{t,1},...f^{\text{ser}}_{t,n}]^{\tt T}$;
 \item MEC server control policy: $\pi_{{\text{ser}}}(\bm q_{t})= \bm f^{\text{ser}}_{t}$;
 \item MEC server immediate cost: $c_{a_{\pi_{\text{ser}}}}(\bm q_t)\triangleq w_1q_{t}^{\text{ser}}+w_2E_t^{\text{ser}}$.
 \end{itemize}
For each WD $k\in{\cal K}$, it follows that
 \begin{itemize}
 \item WD $k$'s local state space: ${\cal S}^{\text{wd}}_{k} =\{\bm s^{\text{wd}}_{k,0}, \bm s^{\text{wd}}_{k,1},...\}$, where $ \bm s^{\text{wd}}_{k,t} \triangleq [q_{k,t}^{\text{wd}},h_{k,t},{\beta}_{k,t,1},...,{\beta}_{k,t,n}]^{\tt T}
     \in\mathbb{C}^{(n+2)\times 1}$;
 \item WD $k$'s local action space: ${\cal A}_k^{\text{wd}}=\{p_{k,0},p_{k,1},...\}$; 
 \item WD $k$'s local control policy: $\pi_{{\text{wd}}.k}(\bm s^{\text{wd}}_{k,t})=p_{k,t}$;
 \item WD $k$'s local immediate cost: $c_{a_{\pi_{{\text{wd}}.k}}}(\bm s^{\text{wd}}_{k,t}) \triangleq w_3Q^{\text{wd}}_{k,t}+w_4E_{k,t}$.
 \end{itemize}
 It is worth noting that the WD $k$'s local action variables should include the local CPU rate vector $\bm f^{\text{wd}}_{k,t}=[f^{\text{wd}}_{k,t,1},...,f^{\text{wd}}_{k,t,n}]^{\tt T}$ in state $\bm s^{\text{wd}}_{k,t}$. Nonetheless, as shown in \eqref{eq.tx-power}, the WD $k$'s CPU rate vector $\bm f^{\text{wd}}_{k,t}$ can be readily determined by the transmit power $p_{k,t}$ in state $\bm s^{\text{wd}}_{k,t}$. Therefore, we herein only specify the transmit power $p_{k,t}$ as the local action variable.

 Note that the cost function $c_{a_{\pi}}(s_t)$ in \eqref{eq.c_a_pi} is of an additive structure. Therefore, we can linearly decompose the cost function $c_{a_{\pi}}(s_t)$ into the $(K+1)$ terms, which are respectively associated with the $K$ WDs and the MEC server. Specifically, we have the cost function decomposition for the MDP (P1) given by
 \begin{align}\label{eq.c-decompose}
 c_{a_{\pi}}(s_t) &= \underbrace{w_1 q_{t}^{\text{ser}} + w_2 E_t^{\text{ser}}}_{ = c_{a_{\pi_{\text{ser}}}}(\bm q_t)} + \sum_{k=1}^K \underbrace{\Big(w_3 q_{k,t}^{\text{wd}} + w_4 E_{k,t}\Big)}_{= c_{a_{\pi_{{\text{wd}}.k}}}(\bm s^{\text{wd}}_{k,t})} \notag \\
 &=
 c_{a_{\pi_{\text{ser}}}}(\bm q_t) + \sum_{k=1}^K c_{a_{\pi_{{\text{wd}}.k}}}(\bm s^{\text{wd}}_{k,t}).
 \end{align}
 With the function decomposition in \eqref{eq.c-decompose} and by exploiting the independency among the $K$ WDs transmit power for task offloading and local computing, we further decompose the centralized MDP problem (P1) as a total of $(K+1)$ subproblems, which are associated with the $K$ WDs and the MEC server, respectively. Specifically, the subproblem associated with WD $k\in{\cal K}$ is given by
 \begin{subequations}\label{eq.prob-wdk}
 \begin{align}
 &({\rm P2.WD}.k): \notag \\
 &\min_{\pi_{{\text{wd}}.k}} J^{\pi_{{\text{wd}}.k}}\triangleq \lim_{T\rightarrow \infty}\mathbb{E}^{\pi_{{\text{wd}}.k}}\Big\{ \sum_{t=0}^T\gamma^t c_{a_{\pi_{{\text{wd}}.k}}}(\bm s^{\text{wd}}_{k,t}) \Big\} \\
 &~~{\rm s. t.}~p_{k.t} =
   \frac{\Gamma}{h_{k,t}}\Big(2^{\frac{D_{k,t}(\bm f_{k,t}^{\text{wd}})}{\tau B_k}}-1 \Big),~\forall t\\
 &~~~~~~~0\leq f^{\text{wd}}_{k,t,i}\leq f^{\text{wd}}_{k,\max},~i=1,...,n, \forall t,
 \end{align}
 \end{subequations}
 where the expectation in (\ref{eq.prob-wdk}a) is taken over different decision makings following the control policy $\pi_{{\text{wd}}.k}$ of WD $k$. For subproblem (P2.WD.$k$), the action-value function $Q^{\pi_{{\text{wd}}.k}}(\bm s^{\text{wd}}_k,p_{k})$ of taking action $p_k$ in state $\bm s^{\text{wd}}_k$ and thereafter taking the policy $\pi_{{\text{wd}}.k}$ is given by
 \begin{align}
 &Q^{\pi_{{\text{wd}}.k}}(\bm s^{\text{wd}}_k,p_k) =\notag \\
 & \lim_{T\rightarrow \infty}\mathbb{E}^{\pi_{{\text{wd}}.k}}\Big\{
 \sum_{t=0}^T\gamma^t c_{a_{\pi_{{\text{wd}}.k}}}(s^{\text{wd}}_{k,t}) \Big| \bm s^{\text{wd}}_{k,0} =\bm s^{\text{wd}}_k,p_{k,0}=p_k
 \Big\}.
 \end{align}

 In addition, the subproblem associated with the MEC server is given by
 \begin{subequations}\label{eq.prob-server}
 \begin{align}
 &({\rm P2.Server}): \notag \\
 &\min_{\pi_{\text{ser}}}~ J^{\pi_{\text{ser}}} \triangleq \lim_{T\rightarrow \infty}\mathbb{E}^{\pi_{\text{ser}}}\Big\{ \sum_{t=0}^T\gamma^t c_{a_{\pi_{\text{ser}}}}(\bm q_t)\Big\} \\
 &~~{\rm s. t.}~0\leq f^{\text{ser}}_{t,i}\leq f^{\text{ser}}_{\max},~i=1,...,n,\forall t,
 \end{align}
 \end{subequations}
 where the expectation in (\ref{eq.prob-server}a) is taken over different decision makings following the control policy $\pi_{\text{ser}}$ of the MEC server. Accordingly, the action-value function $Q^{\pi_{\text{ser}}}(\bm q,\bm f^{\text{ser}})$ associated with subproblem (P2.Server) is given by
 \begin{align}
 &Q^{\pi_{\text{ser}}}(\bm q,\bm f^{\text{ser}}) =\notag \\
 & \lim_{T\rightarrow \infty}\mathbb{E}^{\pi_{\text{ser}}}\left\{
 \sum_{t=0}^T\gamma^t c_{a_{\pi_{{\text{ser}}}}}(\bm q_{t}) \Big| \bm q_{0} =\bm q,\bm f^{\text{ser}}_{0}=\bm f^{\text{ser}}
 \right\}.
 \end{align}

 The decomposition of problem (P1) into subproblems (P2.WD.$k$) and (P2.Server) is always guaranteed to achieve the optimal control policy $\pi^*$ of problem (P1). This is formally stated in the following theorem.

 \begin{theorem}[Equivalent Decomposition of (P1)]\label{theorem.equivalent}
 By solving subproblems (P2.WD.$k$) and (P2.Server) with $k\in{\cal K}$, one can always achieve the optimal expected long-term discounted performance of problem (P1). In other words, it holds that $\pi^*=(\pi^*_{{\text{wd}},1},...,\pi^*_{{\text{wd}},K},\pi^*_{\text{ser}})$ is the optimal solution to (P1).
 \end{theorem}
 \begin{IEEEproof}
 See Appendix~\ref{theorem.proof-equivalent}.
 \end{IEEEproof}

\subsection{Decentralized Bellman Optimality Equations for (P2.WD.$k$) and (P2.Server)}
 For each WD $k\in{\cal K}$, the random state process $\{\bm s^{\text{wd}}_{k,t}\}$ is a controlled Markov process with the transition kernel given by
 \begin{align}\label{eq.tran-kernel-wd}
 & {\rm Pr}(\bm s^{\text{wd}}_{k,t+1}| \bm s^{\text{wd}}_{k,t},a_{\pi_{{\text{wd}}.k}}(\bm s^{\text{wd}}_{k,t}))\notag \\
 &~~={\rm Pr}((h_{k,t+1}, \{\beta_{k,t+1,i},\forall i\},q^{\text{wd}}_{k,t}) | \bm s^{\text{wd}}_{k,t},p_{k,t}) \notag \\
 &~~={\rm Pr}(h_{k,t+1})\prod_{i=1}^n{\rm Pr}(\beta_{k,t+1,i}){\rm Pr}(q^{\text{wd}}_{k,t+1}|\bm s^{\text{wd}}_{k,t},p_{k,t}).
 \end{align}

 Similarly, for the MEC server, the random state process $\{\bm q_t\}$ is a controlled Markov process with the transition kernel given by
 \begin{align}\label{eq.tran-kernel-server}
 &{\rm Pr}(\bm q_{t+1} | \bm q_{t},a_{\pi_{\text{ser}}}(\bm q_{t})) \notag \\
 &~~~~= {\rm Pr}((q^{\text{ser}}_{t+1},q^{\text{wd}}_{1,t+1},...,q^{\text{wd}}_{K,t+1}) | \bm q_{t},\bm f^{\text{ser}}_{t}) \notag \\
 &~~~~= \prod_{k=1}^K{\rm Pr}(q^{\text{wd}}_{k,t+1} | q^{\text{wd}}_{k,t}){\rm Pr}(q^{\text{ser}}_{t+1}| \bm q_{t},\bm f^{\text{ser}}_{t}) ).
 \end{align}

 Based on \eqref{eq.tran-kernel-wd} and \eqref{eq.tran-kernel-server}, subproblems (P2.WD.$k$) and (P2.Server) belong to a class of infinite-horizon MDP problems\cite{Sutton2018}. Therefore, the optimal solutions of sub-MDPs (P2.WD.$k$) and (P2.Server) can be respectively characterized by their Bellman equations. We summarize the decentralized Bellman optimality equations for the optimal action-value functions in the two lemmas as follows.

 \begin{lemma}[Optimality Conditions of sub-MDP~(P2.WD.$k$)]\label{lemma.Bellman-eq-wdk}
 Let $\pi_{{\text{wd}}.k}^* = \argmin_{\pi_{{\text{wd}}.k}}~Q^{\pi_{{\text{wd}}.k}}(\bm s^{\text{wd}}_{k},p_{k})$ and we denote by $Q^*(\bm s^{\text{wd}}_{k},p_{k})=Q^{\pi_{{\text{wd}}.k}^*}(\bm s^{\text{wd}}_{k},p_{k})$ the optimal action-value function. Then, at the optimality of sub-MDP (P2.WD.$k$), the Bellman optimality equation for the action-value function $Q^*(\bm s^{\text{wd}}_{k}, p_{k})$ is given by
 \begin{align}\label{eq.Bellman-eq-wdk}
 & Q^*(\bm s^{\text{wd}}_{k},p_{k}) = \mathbb{E}\Big\{c(\bm s^{\text{wd}}_k,p_k) \notag \\
 &~~~~ + \gamma \min_{p^{\prime}_k} Q^{*}(\bm s^{\text{wd}}_{k,t+1},p^{\prime}_k) \Big| \bm s^{\text{wd}}_{k,t} =\bm s^{\text{wd}}_{k},p_{k,t}=p_{k}\Big\},
 \end{align}
 where $c(\bm s^{\text{wd}}_k,p_{k}) \triangleq c_{a_{\pi_{{\text{wd}}.k}}}(\bm s^{\text{wd}}_k)$, and the expectation in \eqref{eq.Bellman-eq-wdk} is taken over the state transition probabilities of $\{\bm s^{\text{wd}}_{k}\}$.
 \end{lemma}

 \begin{lemma}[Optimality Conditions of  sub-MDP~(P2.Server)]\label{lemma.Bellman-eq-server}
 Let $\pi_{{\text{ser}}}^* = \argmin_{\pi_{{\text{ser}}}}~Q^{\pi_{{\text{ser}}}}(\bm q, \bm f^{\text{ser}})$ and denote by $Q^*(\bm q, \bm f^{\text{ser}})=Q^{\pi_{{\text{ser}}}^*}(\bm q, \bm f^{\text{ser}})$ the optimal action-value function. Then, at the optimality of sub-MDP (P2.Server), the Bellman optimality equation for the action-value function $Q^*(\bm q,\bm f^{\text{ser}})$ is given by
 \begin{align}\label{eq.Bellman-eq-server}
 & Q^*(\bm q, \bm f^{\text{ser}}) = \mathbb{E}\Big\{ c(\bm q,\bm f^{\text{ser}}) \notag \\
 &~~~~ + \gamma \min_{\bm f^{{\text{ser}}\prime}} Q^{*}(\bm q^{\text{wd}}_{t+1},\bm f^{{\text{ser}}\prime}) \Big| \bm q_{t} =\bm q,\bm f^{\text{ser}}_{t}=\bm f^{\text{ser}}\Big\},
 \end{align}
 where $c(\bm q,\bm f^{\text{ser}}) \triangleq c_{a_{\pi_{{\text{ser}}}}}(\bm q)$, and the expectation in \eqref{eq.Bellman-eq-server} is taken over the state transition probabilities of $\{\bm q_t \}$.
 \end{lemma}

 By solving the Bellman optimality equations in Lemmas~\ref{lemma.Bellman-eq-wdk} and~\ref{lemma.Bellman-eq-server}, we can obtain the optimal solution $\pi^*_{{\text{wd}}.k}$ and $\pi^*_{\pi_{\text{ser}}}$ for the sub-MDP (P2.WD.$k$) and (P2.Server), respectively. Compared to the centralized Bellman optimality equation for the MDP (P1) in Lemma~\ref{lemma.P1-Bellman}, the computational complexity and the signaling overhead both have been significantly reduced in solving the decentralized Bellman optimality equations~\eqref{eq.Bellman-eq-wdk} and~\eqref{eq.Bellman-eq-server} due to the decomposition of the state space. It is worth noting that each WD $k\in{\cal K}$ can individually make its decision based on the local WD-QSI and local CSI by solving the sub-MDP (P2.WD.$k$), and the MEC solves the sub-MDP (P2.Server) without requiring the prior knowledge of the random task arrivals or the global CSI. Nonetheless, due to the continuous state/action space in the sub-MDPs (P2.WD.$k$) and (P2.Server), it is still highly challenging to (numerically) obtain the desired control policies $\pi^*_{{\text{wd}}.k}$ and $\pi^*_{\pi_{\text{ser}}}$ based on \eqref{eq.Bellman-eq-wdk} and \eqref{eq.Bellman-eq-server}. Alternatively, we next develop a new parametric online Q-learning algorithm to solve the sub-MDPs (P2.WD.$k$) and (P2.Server).


\section{Decentralized Solution based on the Parametric Online Q-Learning Algorithm}
 In this section, we first approximate the value functions of sub-MDPs (P2.WD.$k$) and (P2.Server) in a parametric form, and then develop a parametric online Q-learning algorithm to learn the parameter set and obtain the control action for each sub-MDP at a WD or the MEC server.

\subsection{Parametric Q-Learning Solution for Sub-MDP (P2.WD.$k$)}
\begin{figure}
 \centering
 \includegraphics[width = 3.5in]{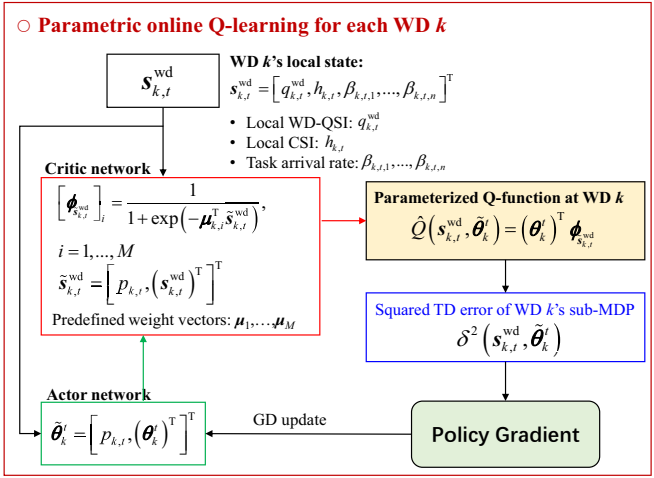}
 \caption{The implementation diagram of parametric online Q-learning for solving the sub-MDP (P2.WD.$k$) at WD $k$, where $k\in{\cal K}$.} \label{fig.network-WD}
\end{figure}

 In this subsection, we first linearly approximate the action-value function $Q^{\pi_{{\text{wd}}.k}}(\bm s_{k,t}^{\text{wd}},p_{k,t})$ of sub-MDP (P2.WD.$k$) in a parametric form with $k\in{\cal K}$, and then develop the parametric online Q-learning algorithm as illustrated in Fig.~\ref{fig.network-WD}.

\subsubsection{Parametric Approximation of the Action-Value Function in (P2.WD.$k$)}
 For the sub-MDP (P2.WD.$k$), we consider using a set of $M$ real-valued parameters $\bm\theta_k\in\mathbb{R}^{M\times 1}$ to linearly approximate the action-value function $Q^{\pi_{{\text{wd}}.k}}(\bm s_{k,t}^{\text{wd}},p_{k,t})$, i.e.,
 \begin{align}\label{eq.linear-value-func-wdk}
 \hat{Q}({\bm s}^{\text{wd}}_{k,t},\tilde{\bm\theta}_k) \triangleq \bm \theta_k^{\tt T} \bm \phi_{\tilde{\bm s}^{\text{wd}}_{k,t}} \approx Q^{\pi_{{\text{wd}}.k}}({\bm s}^{\text{wd}}_{k,t},p_{k,t}),
 \end{align}
 where $\tilde{\bm\theta}_k\triangleq [p_{k,t},(\bm\theta_k)^{\tt T}]^{\tt T}$, and $\bm \phi_{\tilde{\bm s}_{k,t}}\in\mathbb{R}^{M\times 1}$ is a real-valued feature vector characterizing the action-state vector $\tilde{\bm s}^{\text{wd}}_{k,t}\triangleq [p_{k,t},(\bm s^{{\text{wd}}}_{k,t})^{\tt T}]^{\tt T} \in\mathbb{R}^{(n+3)\times 1}$. The $i$th element of the feature vector $\bm \phi_{\tilde{\bm s}^{\text{wd}}_{k,t}}$ is defined as a sigmoid function in the form of
 \begin{align}
 \phi_{\tilde{\bm s}^{\text{wd}}_{k,t}}(\bm \mu_{k,i}^{\tt T}\tilde{\bm s}^{\text{wd}}_{k,t}) = \frac{1}{1+\exp(-\bm \mu_{k,i}^{\tt T}\tilde{\bm s}^{\text{wd}}_{k,t})},~\forall i=1,...,M,
 \end{align}
 where $\bm \mu_{k,i}=[\mu_{k,i}[1],...,\mu_{k,i}[n+3]]^{\tt T} \in\mathbb{R}^{(n+3)\times 1}$ is a predefined real-valued weight vector associated with the action-state vector $\tilde{\bm s}^{\text{wd}}_{k,t}$.

 Note that in the parameterized Q-function $\hat{Q}({\bm s}^{\text{wd}}_{k,t},\tilde{\bm\theta}_k)$ in \eqref{eq.linear-value-func-wdk}, the action variable $p_{k,t}$ (i.e., the transmit power for WD $k$'s task offloading at the $t$th block) and the parameter vector $\bm \theta_k$ and  are free variables.

 \subsubsection{Parametric Online Q-Learning Algorithm}
 We define the temporal-difference (TD) error associated with the Bellman optimality equation \eqref{eq.Bellman-eq-wdk} in Lemma~\ref{lemma.Bellman-eq-wdk} as
 \begin{align}
 \delta(\bm s^{\text{wd}}_{k,t},\tilde{\bm\theta}_k) = r_{k,t} +\gamma \bm \theta_k^{\tt T} \bm \phi_{\tilde{\bm s}^{\text{wd}}_{k,t+1}} - \bm \theta_k^{\tt T} \bm \phi_{\tilde{\bm s}^{\text{wd}}_{k,t}},
 \end{align}
 where $r_{k,t} \triangleq w_3q_{k,t}^{\text{wd}}+\sum_{i=1}^nw_4\tau \xi^{\text{wd}}_k(f^{\text{wd}}_{k,t,i})^3 + w_4\tau p_{k,t}$. The TD error $\delta(\bm s^{\text{wd}}_{k,t},\tilde{\bm\theta}_k)$ represents how closely the parameterized Q-function $\hat{Q}({\bm s}^{\text{wd}}_{k,t},\tilde{\bm \theta}_{k})$ satisfies the Bellman optimality equation. Towards obtaining the optimal vector $\tilde{\bm \theta}^*_k$, we regard the squared TD error as the objective function to be minimized. As such, the error minimization problem of jointly optimizing $p_{k,t}$ and $\bm\theta_k$ is formulated as
 \begin{align}\label{prob.TD-error}
  \tilde{\bm \theta}^*_k= \argmin_{\tilde{\bm\theta}^*_k}~ \delta^2(\bm s^{\text{wd}}_{k,t},\tilde{\bm\theta}_k),~\forall t,
 \end{align}
 where $\tilde{\bm\theta}^*_k=[p^*_{k,t}, ({\bm \theta}^*_k)^{\tt T}]^{\tt T}$.

 Due to the non-convexity relationship between the optimal variable $p_{k,t}$ and the squared TD error function $\delta_{k,t}^2(\bm s^{\text{wd}}_{k,t},\tilde{\bm\theta}_k)$ introduced in the feature function $\phi_{\tilde{\bm s}_{k,t}^{\text{wd}}}$, it is complicated to obtain the closed-form solution $\tilde{\bm \theta}^*_k$ of \eqref{prob.TD-error}. As such, we adopt the celebrated GD method to numerically solve problem \eqref{prob.TD-error}\cite{Sutton2018}. Specifically, the gradient of the objective function $\delta^2(\bm s^{\text{wd}}_{k,t},\tilde{\bm\theta}_k)$ with respect to $p_{k,t}$ is given by
 \begin{align} \label{eq.grad-pkt}
 &\frac{\partial \delta^2(\bm s^{\text{wd}}_{k,t},\tilde{\bm\theta}_k) }{\partial p_{k,t}} = \delta(\bm s^{\text{wd}}_{k,t},\tilde{\bm\theta}_k)\Big[
 w_4\tau - \sum_{i=1}^M\theta_{k}[i]\mu_{k,i}[1] \notag\\
 &~\times \big(\gamma\phi_{\tilde{\bm s}_{k,t+1}^{\text{wd}}}[i](1-\phi_{\tilde{\bm s}_{k,t+1}^{\text{wd}}}[i]) - \phi_{\tilde{\bm s}_{k,t}^{\text{wd}}}[i](1-\phi_{\tilde{\bm s}_{k,t}^{\text{wd}}}[i]) \big) \Big],
 \end{align}
 where $\phi_{\tilde{\bm s}_{k,t+1}^{\text{wd}}}[i] \triangleq \phi_{\tilde{\bm s}_{k,t+1}^{\text{wd}}}(\bm \mu_{k,i}^{\tt T}\tilde{\bm s}_{k,t+1}^{\text{wd}})$ is defined for notational convenience. The partial gradient of $\delta^2(p_{k,t},{\bm\theta}_k)$ with respect to ${\theta}_k[i]$ is given by
 \begin{align}\label{eq.grad-theta}
 \frac{\partial \delta^2(\bm s^{\text{wd}}_{k,t},\tilde{\bm\theta}_k) }{\partial \theta_{k}[i]} &= \delta(\bm s^{\text{wd}}_{k,t},\tilde{\bm\theta}_k)\big( \gamma\phi_{\tilde{\bm s}_{k,t+1}^{\text{wd}}}[i] - \phi_{\tilde{\bm s}_{k,t}^{\text{wd}}}[i] \big),
 \end{align}
 where $i=1,...,M$.

 Based on the GD method, the design variables $(p_{k,t}, {\bm \theta_k})$ of problem \eqref{prob.TD-error} are updated according to
 \begin{subequations}\label{eq.grad-wdk}
 \begin{align}
 p_{k,t+1} &= \Big[ p_{k,t} - \alpha_k^t \times \frac{\partial \delta^2(\bm s^{\text{wd}}_{k,t},\tilde{\bm\theta}^t_k) }{\partial p_{k,t}} \Big]^+\\
 \bm \theta_k^{t+1} &=  \bm \theta_k^t - \alpha_k^t  \delta(\bm s^{\text{wd}}_{k,t},\tilde{\bm\theta}^t_k)\big( \gamma\bm \phi_{\tilde{\bm s}_{k,t+1}^{\text{wd}}} - \bm \phi_{\tilde{\bm s}_{k,t}^{\text{wd}}}\big),
 \end{align}
 \end{subequations}
 where $\tilde{\bm\theta}^t_k=[p_{k,t},(\bm\theta^t_k)^{\tt T}]^{\tt T}$, and $[x]^+\triangleq \max(x,0)$ guarantees the transmit power $p_{k,t+1}$ to be feasible for the sub-MDP (P2.WD.$k$), and $\alpha_k^t$ is a sequence of positive step-size parameters. We summarize the parametric online Q-learning algorithm for obtaining a stationary solution $(p^{**}_{k,t},{\bm \theta}_k^{**})$ for the sub-MDP (P2.WD.$k$) in Algorithm~1.

 With $(p^{**}_{k,t},{\bm \theta}_k^{**})$ obtained, we can obtain the WD $k$'s local CPU rate vector $\bm f_{k,t}^{\text{wd}**}$ by replacing $p_{k,t}$ with $p^{**}_{k,t}$ in (\ref{eq.prob-wdk}b), i.e., $p^{**}_{k.t} =\frac{\Gamma}{h_{k,t}}(2^{\frac{D_{k,t}(\bm f_{k,t}^{{\text{wd}}**})}{\tau B_k}}-1)$. Now, we have completed the parametric online Q-learning algorithm for solving the sub-MDP (P2.WD.$k$).

\begin{table}[htp]
\begin{center}
\caption{Algorithm~1: Parametric Q-Learning for Solving (P2.WD.$k$)}
\hrule
\begin{itemize}
 \item[a)] {\bf Initialization:} Set $t=0$, $\epsilon>0$. Start the system with initial state $\bm s^{\text{wd}}_{k,0}=[q^{\text{wd}}_{k,0},h_{k,0},\beta_{k,0,1},...,
     \beta_{k,0,n}]^{\tt T}$ and transmit power $p_{k,0}$. The policy parameter vector $\bm \theta_k^{0}$ and the step size $\alpha_k^{0}$ are initiated.
 \item[b)] {\bf While} $\frac{\|\bm \theta^{t+1}_{k} - \bm \theta^{t}_k\|}{\|\bm \theta^{t}_k\|}>\epsilon$ {\bf do}
 \begin{itemize}
 \item Compute the gradient $\frac{\partial \delta^2(\bm s^{\text{wd}}_{k,t},\tilde{\bm\theta}^t_k)}{\partial p_{k,t}}$ based on \eqref{eq.grad-pkt};
 \item Compute the gradient $\frac{\partial \delta^2(\bm s^{\text{wd}}_{k,t},\tilde{\bm\theta}^t_k)}{\partial \theta^t_k[i]}$ based on \eqref{eq.grad-theta} for $i=1,...,M$;
 \item Update $(p_{k,t},{\bm \theta}^t_k)$ according to \eqref{eq.grad-wdk};
 \item Set $t \gets t+1$;
    \end{itemize}
\item[c)] {\bf End while}
\item[d)] {\bf Output:} The local-optimal vector $(p_{k,t}^{**}, {\bm\theta}^{**}_k) = (p_{k,t},{\bm \theta}_k^{t})$, the local-optimal transmit power $p^{**}_{k,t}$, and the local-optimal WD $k$'s CPU rate vector $\bm f^{{\text{wd}}**}_{k,t}$ are generated by replacing $p_{k,t}$ with $p_{k,t}^{**}$ in (\ref{eq.prob-wdk}b).
\end{itemize}
\hrule
\end{center}
\end{table}

\subsection{Parametric Q-learning Solution for Sub-MDP (P2.Server)}
\begin{figure}
 \centering
 \includegraphics[width = 3.5in]{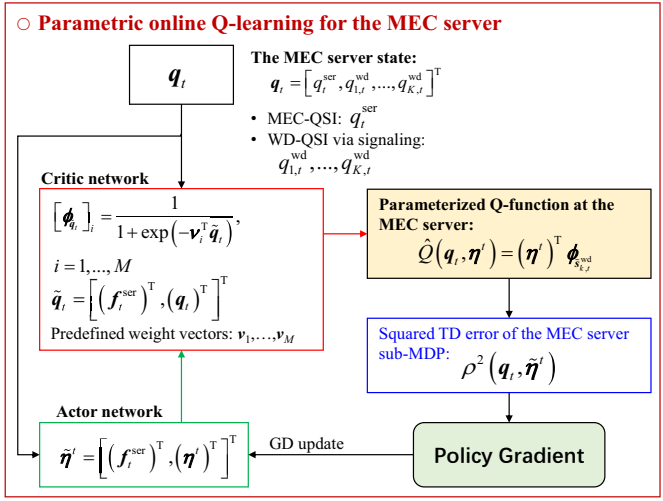}
 \caption{The implementation diagram of parametric online Q-learning for solving the sub-MDP (P2.Server) at the MEC server.} \label{fig.network-server}
\end{figure}

 In this subsection, we first linearly approximate the action-value function $Q^{\pi_{\text{ser}}}(\bm q_t,\bm f^{\text{ser}}_t)$ of the sub-MDP (P2.Server) in a parametric form, and then develop the parametric online Q-learning solution as illustrated in Fig.~\ref{fig.network-server}.

\subsubsection{Parametric Approximation of the Action-Value Function in (P2.Server)}
 Consider a linear parametrization of the value function $Q^{\pi_{\text{ser}}}(\bm q_t,\bm f_t^{\text{ser}})$ of the sub-MDP (P2.Server) as
 \begin{align}\label{eq.linear-value-func-server}
 \hat{Q}({\bm q}_t,\tilde{\bm \eta}) \triangleq \bm \eta^{\tt T} \bm \phi_{\tilde{\bm q}_t} \approx Q^{\pi_{\text{ser}}}(\bm q_{t},\bm f^{\text{ser}}_t),
 \end{align}
 where $\tilde{\bm \eta}\triangleq [({\bm f}_t^{\text{ser}})^{\tt T}, \bm\eta^{\tt T}]^{\tt T}$ with $\bm \eta=[\eta[1],...,\eta[M]]^{\tt T}\in\mathbb{R}^{M\times 1}$ denoting the real-valued parameter vector to be learned, $\tilde{\bm q}_t = [(\bm f^{\text{ser}}_{t})^{\tt T},\bm q^{\tt T}_t]^{\tt T}\in\mathbb{R}^{(n+K+1)\times 1}$ denotes the action-state vector, and $\bm \phi_{\tilde{\bm q}_t}\in\mathbb{R}^{M\times 1}$ is the real-valued feature vector characterizing the MEC server's action-state vector $\tilde{\bm q}_t$. Again, the $i$th element of the feature vector $\bm \phi_{\tilde{\bm q}_t}$ is defined as the sigmoid function, i.e.,
 \begin{align}
 \phi_{\tilde{\bm q}_{t}}(\bm \nu_{i}^{\tt T}\tilde{\bm q}_{t}) = \frac{1}{1+\exp(-\bm \nu_{i}^{\tt T}\tilde{\bm q}_{t})},~\forall i=1,...,M,
 \end{align}
 where $\bm \nu_{i}=[\nu_{i}[1],...,\nu_{i}[n+K+1]]^{\tt T} \in\mathbb{R}^{(n+K+1)\times 1}$ is a predefined real-valued weight vector associated with the action-state vector $\tilde{\bm q}_{t}$. In the linearly parameterized Q-function $\hat{Q}({\bm q}_{t},\tilde{\bm \eta})$ in \eqref{eq.linear-value-func-server}, the MEC server's CPU rate vector $\bm f^{\text{ser}}_{t}$ and the parameter vector $\bm\eta$ are the free design variables.

\subsubsection{Parametric Online Q-Learning Algorithm}
 By substituting the parameterized Q-function $ \hat{Q}({\bm q}_{t},\tilde{\bm \eta})$ and removing the expectation in the Bellman optimality equation \eqref{eq.Bellman-eq-server}, we obtain the TD error as a function of the vector $\tilde{\bm \eta}=[(\bm f^{\text{ser}}_t)^{\tt T},({\bm \eta})^{\tt T}]^{\tt T}$, i.e.,
 \begin{align}
 \rho(\bm q_t, \tilde{\bm \eta}) = r_{t} +\gamma \bm \eta^{\tt T} \bm \phi_{\tilde{\bm q}_{t+1}} - \bm \eta^{\tt T} \bm \phi_{\tilde{\bm q}_{t}},
 \end{align}
 where $\bm q_t$ is the state realization of the sub-MDP~(P2.WD.$k$) including the MEC-QSI $q_t^{\text{ser}}$ and WD-QSI $\{q_{1,t}^{\rm wd},...,q_{K,t}^{\rm wd}\}$, and
  $r_{t} \triangleq w_1q_{t}^{\text{ser}}+\sum_{i=1}^nw_2\tau \xi^{\text{ser}}(f^{\text{ser}}_{t,i})^3$. The TD error $\rho(\bm q_t, \tilde{\bm \eta})$ indicates how closely the parameterized Q-function $\hat{Q}({\bm q}_{t},\tilde{\bm \eta})$ satisfies the Bellman optimality equation \eqref{eq.Bellman-eq-server} in Lemma~\ref{lemma.Bellman-eq-server}. Similar to the online Q-learning procedure of the vector $(p_{k,t},{\bm \theta}_k)$ for the sub-MDP (P2.WD.$k$), we consider the squared TD error (i.e., $\rho^2(\bm q_t, \tilde{\bm \eta})$) as the objective function to be minimized by optimizing $\tilde{\bm \eta}$. Therefore, the squared TD error minimization problem of interest is formulated as
 \begin{align}\label{prob.TD-error-server}
 \tilde{\bm \eta}^* = \argmin_{\tilde{\bm \eta}^*}~ \rho^2(\bm q_t, \tilde{\bm \eta}),~\forall t,
 \end{align}
 where $\tilde{\bm \eta}^*=[(\bm f^{\text{ser}*}_t)^{\tt T}, ({\bm \eta}^*)^{\tt T}]^{\tt T}$.

 Again, it is highly challenging to analytically solve \eqref{prob.TD-error-server} due to the non-convexity of the feature function $\phi_{\tilde{\bm q}_{t}}(\bm \nu_{i}^{\tt T}\tilde{\bm q}_{t})$ with respect to the vector $\tilde{\bm \eta}$, where the MEC server's control action $\bm f^{\text{ser}}_t$ acts as a partial design variable vector in \eqref{prob.TD-error-server}. As such, we employ the GD method to find a stationary solution to problem \eqref{prob.TD-error-server}. The gradient of the squared TD error $\rho^2(\bm q_t,\tilde{\bm\eta})$ with respect to the MEC server's CPU rate $f^{\text{ser}}_{t,i} = \tilde{\eta}[i]$ at the $i$th slot of the $t$th block is given by
 \begin{align}\label{eq.grad-cpu-rate-server}
 & \frac{\partial \rho^2(\bm q_t,\tilde{\bm\eta})}{\partial f^{\text{ser}}_{t,i}} = \rho(\bm q_t, \tilde{\bm\eta})\Big[
 3w_2\tau\xi^{\text{ser}}(f_{t,i}^{\text{ser}})^2 - \sum_{j=1}^{M}\eta[j]\nu_j[i] \notag \\
 &~~~ \times \big( \gamma \phi_{\tilde{\bm q}_{t+1}}[j](1-\phi_{\tilde{\bm q}_{t+1}}[j]) - \phi_{\tilde{\bm q}_{t}}[j](1-\phi_{\tilde{\bm q}_{t}}[j]) \big) \Big],
 \end{align}
 where $i=1,...,n$. The gradient of the squared TD error function $\rho^2(\bm q_t,\tilde{\bm\eta})$ with respect to the parameter vector $\bm \eta$ is given by
 \begin{align}\label{eq.grad-eta}
 \frac{\partial \rho^2(\bm q_t,\tilde{\bm\eta}) }{\partial \bm \eta} &= \rho(\bm q_t, \tilde{\bm\eta})\big(\gamma\bm \phi_{\tilde{\bm q}_{t+1}} - \bm \phi_{\tilde{\bm q}_{t}} \big).
 \end{align}

 Based on the GD method, the design variables $(\bm f^{\text{ser}}_t,{\bm \eta})$ of problem \eqref{prob.TD-error-server} are updated according to
 \begin{subequations}\label{eq.grad-server}
 \begin{align}
 f^{\text{ser}}_{t+1,i} &=  \Big[f^{\text{ser}}_{t,i} - \alpha^t \times \frac{\partial \rho^2(\bm q_t, \tilde{\bm\eta}^t)}{\partial f^{\text{ser}}_{t,i}}\Big]^{f_{\max}^{\text{ser}}}_0,\\
 \bm \eta^{t+1} &=  \bm \eta^t - \alpha^t  \rho(\bm q_t, \tilde{\bm\eta}^t)\big( \gamma\bm \phi_{\tilde{\bm q}_{t+1}} - \bm \phi_{\tilde{\bm q}_{t}}\big),
 \end{align}
 \end{subequations}
 where $i=1,...,n$, $\tilde{\bm\eta}^t=[(\bm f^{\text{ser}}_{t})^{\tt T},(\bm\eta^t)^{\tt T}]^{\tt T}$, and $\alpha^t$ is a sequence of positive step-size parameters. The parametric online Q-learning method for obtaining a stationary solution $(\bm f^{\text{ser}**}_t, {\bm \eta}^{**})$ for the sub-MDP (P2.Server) is presented in Algorithm~2.

 Now, we have completed the parametric online Q-learning algorithm for the sub-MDP (P2.Server).

\begin{table}[htp]
\begin{center}
\caption{Algorithm~2: Parametric Q-Learning for Solving (P2.Server)}
\hrule
\begin{itemize}
 \item[a)] {\bf Initialization:} Set $t=0$, $\epsilon>0$. Start the system with initial state $\bm q_{0}=[q^{\text{ser}}_0,q^{\text{wd}}_{k,0},...,q^{\text{wd}}_{K,0}]^{\tt T}$ and transmit power $\bm f^{\text{ser}}_{0}=[f^{\text{ser}}_{0,1},...,f^{\text{ser}}_{0,n}]^{\tt T}$. The policy parameter vector $\bm \eta^{0}$ and the step-size $\alpha^{0}$ are initiated.
 \item[b)] {\bf While} $\frac{\|\bm \eta^{t+1} - \bm \eta^{t}\|}{\|\bm \eta^{t}\|}>\epsilon$ {\bf do}
 \begin{itemize}
 \item Compute the gradient $\frac{\partial \rho^2(\bm q_t,\tilde{\bm\eta}^t)}{\partial f^{\text{ser}}_{t,i}}$ based on  \eqref{eq.grad-cpu-rate-server};
 \item Compute the gradient $\frac{\partial \rho^2(\bm q_t,\tilde{\bm\eta}^t)}{\partial \bm \eta^t}$ based on  \eqref{eq.grad-eta};
 \item Update $(\bm f^{\text{ser}}_t, {\bm \eta}^t)$ according to \eqref{eq.grad-server};
 \item Set $t \gets t+1$;
    \end{itemize}
\item[c)] {\bf End while}
\item[d)] {\bf Output:} The local-optimal vector ${\bm\eta}^{**} = {\bm \eta}^{t}$ and the local-optimal MEC server's CPU rate vector $\bm f^{{\text{ser}}**}_{t}=\bm f^{{\text{ser}}}_{t}$.
\end{itemize}
\hrule
\end{center}
\end{table}

\begin{proposition}[Convergence Analysis]\label{prop.convergence}
 Let the step-size $\alpha^t_k$ and $\alpha^t$ satisfy
 \begin{subequations}
 \begin{align}
 & \sum_{t=1}^\infty \alpha_k^{t} = \infty,~~\sum_{t=1}^\infty (\alpha_k^{t})^2 < \infty,\forall k\in{\cal K}\\
 & \sum_{t=1}^\infty \alpha^{t} = \infty,~~\sum_{t=1}^\infty (\alpha^{t})^2 < \infty.
\end{align}
\end{subequations}
Then, the solution $(p_{k,t}^{**},{\bm \theta}_k^{**})$ and $(\bm f^{\text{ser}**}_{t},{\bm \eta}^{**})$ obtained by the parametric online Q-learning method (i.e., Algorithms 1 and 2) converges to the neighborhood of a stationary point of problems \eqref{prob.TD-error} and \eqref{prob.TD-error-server}, respectively, in the sense that:
\begin{align}\label{eq.conv-WD}
\liminf_{t\geq0}\mathbb{E}\left[\left\Vert \nabla_{\tilde{\bm\theta}_{k}^{t}}\delta^{2}(\bm s_{k,t}^{\text{wd}},\tilde{\bm\theta}^t_k)\right\Vert_{2}^{2}\right]
\leq\frac{\mathbb{E}\left[\delta^{2}(\bm s^{\text{wd}}_{k,2},\tilde{\bm\theta}_{k}^{1})\right]}{\mathrm{E_{1}}\left(\ln\frac{1}{\gamma}\right)}
\end{align}
and
\begin{align}\label{eq.conv-MEC}
 \liminf_{t\geq0}\mathbb{E}\left[\left\Vert\nabla_{\tilde{\bm\eta}^{t}}\rho^{2}(\bm q_t,\tilde{\bm \eta}^t)\right\Vert_{2}^{2}\right]
 \leq\frac{\mathbb{E}\left[\rho^{2}(\bm q_{2},\tilde{\bm\eta}^{1})\right]}{\mathrm{E_{1}}\left(\ln\frac{1}{\gamma}\right)},
\end{align}
where $\tilde{\bm\theta}^t_k\triangleq [p_{k,t},(\bm \theta_k^t)^{\tt T}]^{\tt T}$, $\tilde{\bm\eta}^t\triangleq [(\bm f^{\text{ser}}_{t})^{\tt T},(\bm \eta^t)^{\tt T}]^{\tt T}$, and $\mathbb{E}[\delta^2(\bm s^{\text{wd}}_{k,2},\tilde{\bm\theta}_{k}^{1})]$ and $\mathbb{E}[\rho^2(\bm q_{2},\tilde{\bm\eta}^{1})]$ are the bounded constants representing the squared TD error of the first time block at the $k$th WD and the MEC server, respectively, and $\mathrm{E_{1}}(x)=\int_{x}^{\infty}\frac{e^{-t}}{t}dt$ is the exponential integral function.
\end{proposition}

\begin{IEEEproof}
See Appendix~\ref{proof-prop-convergence}.
\end{IEEEproof}

\begin{remark}
 From \eqref{eq.conv-WD} and \eqref{eq.conv-MEC} in Proposition~\ref{prop.convergence}, we observe that the exponential integral function $\mathrm{E_{1}}\left(\ln\frac{1}{\gamma}\right)$ approaches infinity as the discount factor $\gamma$ tends to 1. Accordingly, we have the results of $\lim_{\gamma\rightarrow 1}\liminf_{t\geq0}\mathbb{E}\left[\left\Vert \nabla_{\tilde{\bm\theta}_{k}^{t}}\delta^{2}(\bm s_{k,t}^{\text{wd}},\tilde{\bm\theta}^t_k)\right\Vert_{2}^{2}\right]=0$ and $\lim_{\gamma\rightarrow 1}\liminf_{t\geq0}\mathbb{E}\left[\left\Vert\nabla_{\tilde{\bm\eta}^{t}}\rho^{2}(\bm q_t,\tilde{\bm \eta}^t)\right\Vert_{2}^{2}\right]=0$. Note that when the discount factor $\gamma$ approaches 1, the discounted-cost MDP problem (P1) will become an average-cost MDP problem. Therefore, our proposed parametric online Q-learning algorithm can converge to a stationary point under the average cost MDP setup, which coincides with the convergence results in the existing actor-critic algorithms \cite{Actor-Critic03}.
\end{remark}

\section{Numerical Results}
 In this section, we provide numerical results to evaluate the performance of the proposed designs. In the simulations, the system parameters are set as follows unless otherwise stated. The $K$ WDs are randomly and uniformly distributed in the cell with a radius of 200~meters. The pathloss between the BS and WDs is modelled as $30.6+37.6\log_{10}d_k$, where $d_k$ is the corresponding distance for each WD $k$ to the BS, and the standard derivation of shadow fading is 10~dB\cite{Goldsmith-Book}. The noise power at the BS receiver is set to be -174~dBW. The Bernoulli distribution of the DNN task arrival at each WD $k\in{\cal K}$ is set to have the probability $\text{Pr}(\beta_{k,t,i}=1)=b_k=0.4$. The time block is set to include $n=5$ time slots, and the time duration of each slot is $\tau=0.1$ seconds. The workload of each DNN task is set to consist of $W=10^10$ CPU cycles, and the effective capacitance of the CPU architecture for WD $k\in{\cal K}$ and the MEC server is set as $\xi_k^{\text{wd}}=10^{-28}$ Joules and $\xi_k^{\text{ser}}=10^{-29}$ Joules, respectively. For each WD $k\in{\cal K}$, the spectrum bandwidth for task offloading is set as $B_k=1$ MHz and the SNR gap is set as $\Gamma = 1.5$\cite{Goldsmith-Book}. The error tolerance is set as $\epsilon=10^{-3}$ and the parameter dimension is set as $M=30$ for linearly approximating the action-value function in the proposed parametric online Q-learning Algorithm~1 and Algorithm~2. 

For performance comparison, we consider the following three baseline schemes.
\begin{itemize}
\item{\em Binary Offloading Scheme:} In this scheme, the joint task offloading and computation offloading is implemented, where each task is offloaded as an entire entity\cite{Infocom19,DNN-MEC-GC20,EdgeAI-TWC20}. Therefore, each task can only be executed by the WD or the MEC server.
\item{\em Even Resource Allocation Scheme:} In this scheme, the MEC server fairly allocates the computational resources for each WD evenly, i.e., each WD offloads the same amount of task workload to the MEC server\cite{Multiuser-IOTJ21,ACM18}.
\item{\em Random Scheme:} In this scheme, each WD offloads a random amount of the partitioned task workload to the MEC server without considering the competition of resource allocation against other WDs.
\end{itemize}

\begin{figure}
  \centering
  \includegraphics[width = 3.5in]{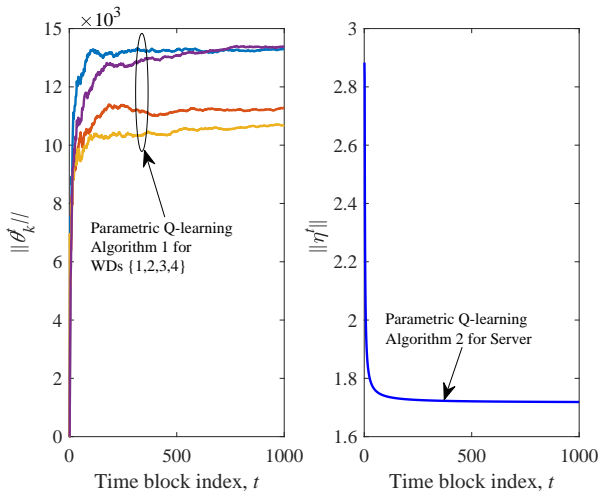}
 \caption{The convergence performance of the proposed parametric Q-learning Algorithm~1 for WDs and Algorithm~2 for the MEC server.} \label{fig.Conv}
\end{figure}

 Figure~\ref{fig.Conv} shows the fast convergence performance of the parametric online Q-learning Algorithm~1 and Algorithm~2, where the number of WDs is $K=4$. It is observed that Algorithm~1 requires around 500 iterations to converge into the desirable solution for the WDs. By contrast, the proposed Algorithm~2 is observed to require around 200 iterations to converge into the desirable solution for the MEC server. This is expected, since the WDs have to additionally handle the task partitioning procedure to adapt to both dynamic task arrivals and wireless offloading channel fluctuation.

\begin{figure}
  \centering
  \includegraphics[width = 3.5in]{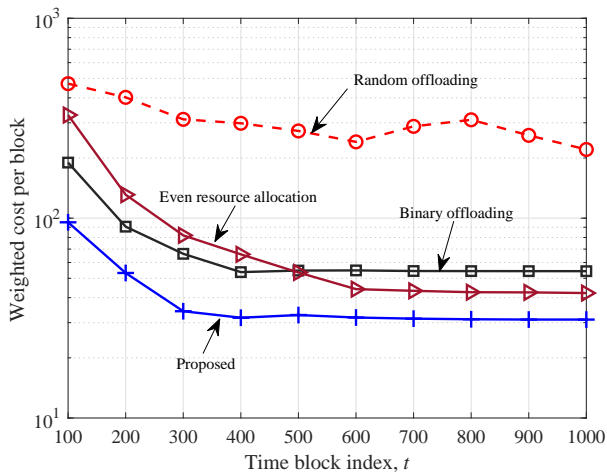}
 \caption{The weighted cost of energy consumption and workload length per time block.} \label{fig.per_slot}
\end{figure}

 Figure~\ref{fig.per_slot} shows the weighted cost of energy consumption and workload length per block, where the number of WDs is $K=20$. It is observed that the cost of the proposed scheme, as well as the baseline binary-offloading and even-resource-allocation schemes, decreases with the increasing of the time block index. This is due to the improvement of the parameter adjustment in the parametric online Q-learning process over time in the three schemes. The proposed scheme is observed to achieve a significant performance gain compared to the three baseline schemes, which is due to the flexible task partitioning capability of the WDs for fully collaboratively exploiting the WD local computation and MEC server resources. The baseline binary-offloading scheme is observed to outperform the even-resource-allocation scheme when the time block index is small (e.g., smaller than 500), but it is not true as the block index $t$ becomes larger. This shows the benefit of reducing the long-term DNN execution cost by leveraging both the local and remote computing capability in the multiuser MEC system. Due to the lack of offloading design optimization, it is also observed that the baseline random-offloading scheme performs inferiorly to the other schemes.

 \begin{figure}
  \centering
  \includegraphics[width = 3.5in]{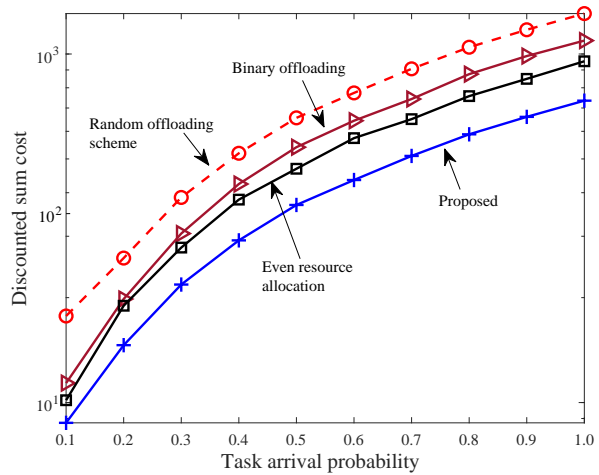}
 \caption{The long-term discounted sum cost versus the task arrival probability at the WDs.} \label{fig.vs_rate}
\end{figure}

 Figure~\ref{fig.vs_rate} shows the long-term discounted sum cost versus the DNN task arrival probability, where the number of WDs is $K=50$. As expected, it is observed that the discounted sum cost of the four schemes increases as the task probability increases. The proposed scheme is observed to outperform the three baseline schemes. This implies the benefit of performing the flexible task partitioning capability and sequential offloading. The baseline binary-offloading scheme is observed to outperform the random-offloading scheme, but it performs inferiorly to the even-resource-allocation scheme in this setup. This indicates the importance of leveraging both the local and remote computing capability for improving the multiuser MEC system performance.

\begin{figure}
  \centering
  \includegraphics[width = 3.5in]{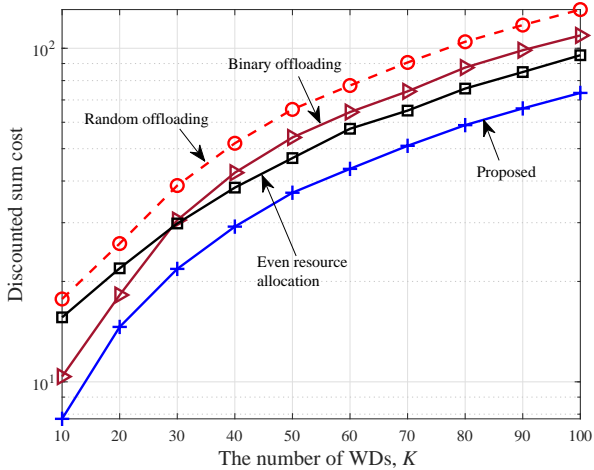}
 \caption{The long-term discounted sum cost versus the number of WDs $K$.} \label{fig.vs_K}
\end{figure}

 Figure~\ref{fig.vs_K} shows the long-term discounted sum cost versus the number of WDs $K$. The discounted sum cost of the four schemes is observed to increase as $K$ increases. Similar to Fig.~\ref{fig.vs_rate}, the proposed scheme is observed to outperform the three baseline schemes. The baseline binary-offloading scheme is observed to outperform the even-resource-allocation scheme with small values of $K$ (e.g., $K\leq 30$), but it performs inferiorly to the even-resource-allocation scheme when $K$ becomes larger in this setup. This indicates the necessity to simultaneously leverage both the local and remote computing capability in the MEC systems with a larger number of WDs. In addition, it is again observed that the baseline random-offloading scheme performs inferiorly to the other schemes.

\section{Conclusion}
 In this paper, we studied the sequential task offloading problem for a multiuser MEC system with dynamic DNN task arrivals and channel fluctuations over time, by jointly optimizing the transmit power for offloading, the partitioning of local DNN tasks, and the local/remote computing rate control. We introduced the WD-QSI and MEC-QSI to capture the dynamic urgency of the tasks at each WD and the MEC server, respectively, and formulated an infinite-horizon MDP problem to minimize the long-term expected discounted system energy consumption and computation delay. To overcome the computational complexity and signaling overhead challenges in solving the centralized Bellman equation, we decomposed the MDP into multiple lower dimensional sub-MDPs, each of which can be associated with a WD or the MEC server. We further developed the parametric online Q-learning algorithm to individually solve each sub-MDP at its associated WD or the MEC server. We proved the convergence result, and provided numerical results to show the superb performance achieved by the proposed decentralized solution over the existing baselines.

\appendix
\subsection{Proof of Theorem~\ref{theorem.equivalent}}\label{theorem.proof-equivalent}
 We prove the equivalent decomposition of problem (P1) from the following two aspects. First, under the fixed control action $\bm f^{\text{ser}}_{t}$ and the MEC-QSI $q_{t}^{\text{ser}}$ of the MEC server at the $t$th block, the immediate cost $(w_1q_{t}^{\text{ser}}+w_2E_t^{\text{ser}})$ is fixed. In this case, problem (P1) reduces to the stochastic optimization of the $K$ WDs' local QSI and CPU rates at the $t$th block. Define
 $\tilde{c}_{a_{\pi}}(s_t) \triangleq c_{a_{\pi}}(s_t) - w_1 q_{t}^{\text{ser}} - w_2 E_t^{\text{ser}}$. Without considering the term $w_1q_{t}^{\text{ser}}+w_2E_t^{\text{ser}}$ in the cost function in \eqref{eq.c_a_pi}, we have
 \begin{align}
 \tilde{c}_{a_{\pi}}(s_t) = \sum_{k=1}^K c_{ a_{\pi_{{\text{wd}}.k}}}(\bm s^{\text{wd}}_{k,t}),
 \end{align}
 where $c_{a_{\pi_{{\text{wd}}.k}}}(\bm s^{\text{wd}}_{k,t})=w_3Q_{k,t}^{\text{wd}} + w_4E_{k,t}$. Note that $c_{a_{\pi_{{\text{wd}}.k}}}(\bm s^{\text{wd}}_{k,t})$ only depends on the local QSI and control action of WD $k$. In addition, the constraints (\ref{eq.v-pi}b--c) of problem (P1) are fully decoupled among the $K$ WDs without loss of optimality. Therefore, under the fixed control action and QSI of the MEC server, problem (P1) can be equivalently decomposed into $K$ subproblems, in which the subproblem to be solved by WD $k$ is expressed as subproblem (P2.WD.$k$).

 Second, after the MEC server collects the knowledge of the WD-QSI and the local computing CPU rates of the $K$ WDs at the $t$th block, the amount of new workload arrival from the $K$ WDs (i.e., $\sum_{k=1}^K (\big[Q^{\text{wd}}_{k,t} - \sum_{i=1}^n f_{k,t,i}^{\text{wd}}\tau\big]^+~{\rm mod}~W)$) at the $t$th block is determined. In this case, problem (P1) reduces to the optimization of minimizing $(w_1q_{t}^{\text{ser}}+w_2E_t^{\text{ser}})$ subject to the constraint (\ref{eq.v-pi}d). Therefore, by considering the long-term discounted performance, the stochastic optimization at the MEC server is expressed as subproblem (P2.Server).

\subsection{Proof of Proposition~\ref{prop.convergence}} \label{proof-prop-convergence}
First, for each WD $k\in{\cal K}$, we prove the convergence of $\tilde{\bm\theta}_{k}^{t}=\left[p_{k,t},\left({\bm \theta}_{k}^{t}\right)^{{\tt T}}\right]^{{\tt T}}$
to a region with a bounded partial derivative, i.e., $\liminf_{t\geq0}\mathbb{E}\left[\left\Vert \nabla_{\tilde{\bm\theta}_{k}^{t}}\delta_{k,t}^{2}\right\Vert _{2}^{2}\right]\leq \text{Bounded-Constant}$.

Denote $\delta(\bm s_{k,t},\tilde{\bm\theta}_{k}^{t})=\delta_{k,t}=r_{k,t}+\gamma\hat{Q}(\bm s^{\text{wd}}_{k,t+1},\tilde{\bm\theta}_{k}^{t})-\hat{Q}(\bm s^{\text{wd}}_{k,t},\tilde{\bm\theta}_{k}^{t})$, where $\hat{Q}(\bm s^{\text{wd}}_{k,t},\tilde{\bm\theta}_{k}^{t})$ is the parameterized Q-function of the $k$th WD. Denote by $L$ the positive constant such that
$\nabla_{\tilde{\bm\theta}_{k}^{t}}\delta_{k,t}^{2}\leq L\mathbf{I}$. Therefore, we have
\begin{align}
 & \mathbb{E}\left[\delta^{2}(\bm s^{\text{wd}}_{k,t+1},\tilde{\bm\theta}_{k}^{t+1})-\delta^{2}(\bm s^{\text{wd}}_{k,t+1},\tilde{\bm\theta}_{k}^{t})\right] \notag \\
& \leq\mathbb{E}\left[-\alpha_{k}^{t}\left(\nabla_{\tilde{\bm\theta}_{k}^{t}}\delta_{k,t}^{2}\right)^{\text{T}}
\nabla_{\tilde{\bm\theta}_{k}^{t}}\delta_{k,t}^{2}+\frac{1}{2}L\left(\alpha_{k}^{t}\right)^{2}\left\Vert \nabla_{\tilde{\bm\theta}_{k}^{t}}\delta_{k,t}^{2}\right\Vert_{2}^{2}\right]\nonumber \\
 & =-\mathbb{E}\left[\tilde{\alpha}_{t}\left\Vert \nabla_{\tilde{\bm\theta}_{k}^{t}}\delta_{k,t}^{2}\right\Vert_{2}^{2}\right],\label{eq: eq-1}
\end{align}
where $\tilde{\alpha}_{k}^{t}=\alpha_{k}^{t}-\frac{1}{2}L(\alpha_{k}^{t})^{2}$ is a positive number and $\alpha_{k}^{t}$ is the positive GD step-size for WD $k$. Furthermore, we note that
\begin{align}\label{eq:eq-2}
&\mathbb{E}\left[\delta^{2}(\bm s^{\text{wd}}_{k,t+1},\tilde{\bm\theta}_{k}^{t+1})-\gamma\delta^{2}(\bm s^{\text{wd}}_{k,t+2},\tilde{\bm\theta}_{k}^{t+1})\right]=\mathbb{E}[r_{k,t+1}].
\end{align}

By combining (\ref{eq: eq-1}) and (\ref{eq:eq-2}), it follows that
\begin{align}
 & \mathbb{E}\left[\delta^{2}(\bm s^{\text{wd}}_{k,t+1},\tilde{\bm\theta}_{k}^{t+1})-\delta^{2}(\bm s^{\text{wd}}_{k,t+1},\tilde{\bm\theta}_{k}^{t})\right]\notag \\
 & \quad -\mathbb{E}\left[\delta^{2}(\bm s^{\text{wd}}_{k,t+1},\tilde{\bm\theta}_{k}^{t+1})-\gamma\delta^{2}(\bm s^{\text{wd}}_{k,t+2}, \tilde{\bm\theta}_{k}^{t+1})\right]\nonumber \\
 & =\mathbb{E}\left[\gamma\delta^{2}(\bm s^{\text{wd}}_{k,t+2},\tilde{\bm\theta}_{k}^{t+1})-\delta^{2}(\bm s^{\text{wd}}_{k,t+1},\tilde{\bm\theta}_{k}^{t})\right] \notag \\
 & \leq-\mathbb{E}\left[\tilde{\alpha}_{k}^{t}
 \left\Vert\nabla_{\tilde{\bm\theta}_{k}^{t}}\delta_{k,t}^{2}\right\Vert_{2}^{2}\right] -\mathbb{E}[r_{k,t+1}].\label{eq:eq-3}
\end{align}
As a result, for any $T\geq 1$, we have
\begin{align}
 & \mathbb{E}\left[\gamma^{T}\delta^{2}(\bm s^{\text{wd}}_{k,T+2},\tilde{\bm\theta}_{k}^{T+1})-\delta^{2}(\bm s^{\text{wd}}_{k,2},\tilde{\bm\theta}_{k}^{1})\right] \notag \\
 &\quad \leq - \sum_{t=1}^{T}\gamma^{t}\left(\tilde{\alpha}_{k}^{t}
 \mathbb{E}\left[\left\Vert \nabla_{\tilde{\bm\theta}_{k}^{t}}\delta_{k,t}^{2}\right\Vert_{2}^{2}\right] +\mathbb{E}[r_{k,t+1}]\right).\label{eq:eq-4}
\end{align}

Re-arranging the terms in (\ref{eq:eq-4}), it follows that
\begin{align}
 & \sum_{t=0}^{T}\gamma^{t}\tilde{\alpha}_{k}^{t}
 \mathbb{E}\left[\left\Vert \nabla_{\tilde{\bm\theta}_{k}^{t}}\delta_{k,t}^{2}\right\Vert_{2}^{2}\right]
 \leq\mathbb{E}\left[\delta^{2}(\bm s^{\text{wd}}_{k,2},\tilde{\bm\theta}_{k}^{1})\right].
\end{align}

Let the GD step-size $\alpha_{k}^{t}$ be chosen such that $\tilde{\alpha}_{k}^{t}=\alpha_{k}^{t}-\frac{1}{2}L(\alpha_{k}^{t})^{2}=\frac{1}{t}$
and take the limit $T\rightarrow\infty$, it follows that
\begin{align}
  \lim_{T\rightarrow\infty}\Big(\sum_{t=1}^{T}\frac{\gamma^{t}}{t}\Big)\liminf_{t\geq0}
  \mathbb{E}\Big[\left\Vert \nabla_{\tilde{\bm\theta}_{k}^{t}}\delta_{k,t}^{2}\right\Vert_{2}^{2}\Big]
  \leq\mathbb{E}\left[\delta^{2}(\bm s^{\text{wd}}_{k,2},\tilde{\bm\theta}_{k}^{1})\right].\label{eq:eq-6}
\end{align}

Furthermore, we note that
\begin{align}
 \lim_{T\rightarrow\infty}\left(\sum_{t=1}^{T}\frac{\gamma^{t}}{t}\right) &\geq\int_{1}^{\infty}\frac{\gamma^{t}}{t}dt\notag \\ &=\mathrm{Ei}\left(\infty\ln\gamma\right)-\mathrm{Ei}\left(\ln\gamma\right)
 = -\mathrm{Ei}\left(\ln\gamma\right)\notag\\
 &=\mathrm{E_{1}}\left(\ln\frac{1}{\gamma}\right)>0,\label{eq:eq-7}
\end{align}
where $\mathrm{Ei}\left(x\right)=\int_{-\infty}^{x}\frac{e^{t}}{t}dt$ is the exponential integral function and $\mathrm{E}_{1}\left(x\right)=\int_{x}^{\infty}\frac{e^{-t}}{t}dt$.

By combining (\ref{eq:eq-6}) and (\ref{eq:eq-7}), it follows that
\begin{align}
 \liminf_{t\geq0}\mathbb{E}\left[\left\Vert \nabla_{\tilde{\bm\theta}_{k}^{t}}\delta_{k,t}^{2}\right\Vert_{2}^{2}\right]
 \leq\frac{\mathbb{E}\left[\delta^{2}(\bm s^{\text{wd}}_{k,2},\tilde{\bm\theta}_{k}^{1})\right]}{\mathrm{E_{1}}\left(\ln\frac{1}{\gamma}\right)},\label{eq:eq-8}
\end{align}
where the right hand side of (\ref{eq:eq-8}) is a bounded constant. Therefore, the proof of the convergence for each WD $k\in{\cal K}$ is completed.

Next, we prove the convergence of $\tilde{\bm\eta}^{t}=\left[(\bm f_{t}^{\text{ser}})^{\text{T}},({\bm \eta}^{t})^{\tt T}\right]^{\tt T}$ to a region with a bounded partial derivative for the MEC server, i.e., $\liminf_{t\geq0}\mathbb{E}\left[\left\Vert\nabla_{\tilde{\bm\eta}^{t}}\rho_{t}^{2}\right\Vert_{2}^{2}\right] \leq \text{Bounded-Constant}$. The proof follows similarly for the convergence of each WD $k$'s case. The key steps are outlined as follows.

Denote $\rho(\bm q_{t},\tilde{\bm\eta}^{t})=\rho_{t}=r_{t}+\gamma\hat{Q}(\bm q_{t+1},\tilde{\bm\eta}^{t}) - \hat{Q}(\bm q_{t},\tilde{\bm\eta}^{t})$, where $\hat{Q}(\bm q_{t},\tilde{\bm\eta}^{t})$ is the parameterized Q-function of the MEC server. Denote by $\tilde{L}$ the positive constant such that $\nabla_{\tilde{\bm\eta}^{t}}\rho_{t}^{2}\leq \tilde{L}\mathbf{I}$. Therefore, it is yielded that
\begin{align}
 \mathbb{E}\left[\rho^{2}(\bm q_{t+1},\tilde{\bm\eta}^{t+1})-\rho^{2}(\bm q_{t+1},\tilde{\bm\eta}^{t})\right] \leq
 -\mathbb{E}\left[\tilde{\alpha}^{t}\left\Vert\nabla_{\tilde{\bm\eta}^{t}}\rho_{t}^{2}\right\Vert_{2}^{2}\right],\label{eq:eq-9}
\end{align}
where $\tilde{\alpha}^t=\alpha^t-\frac{1}{2}\tilde{L}(\alpha^t)^2$ is a positive number and $\alpha^t$ is the positive GD step-size for the MEC server. Furthermore, we have
\begin{align}
 & \mathbb{E}\left[\rho^{2}(\bm q_{t+1},\tilde{\bm\eta}^{t+1}-\gamma\rho^{2}(\bm q_{t+2}, \tilde{\bm\eta}^{t+1})\right]=\mathbb{E}[r_{t+1}].\label{eq:eq-10}
\end{align}

Based on (\ref{eq:eq-9}) and (\ref{eq:eq-10}), it follows that
\begin{align}
 & \mathbb{E}\left[\gamma\rho^{2}(\bm q_{t+2},\tilde{\bm\eta}^{t+1}) - \rho^{2}(\bm q_{t+1},\tilde{\bm\eta}^{t})\right] \notag \\
&\quad \leq-\mathbb{E}\left[\tilde{\alpha}^{t}\left\Vert \nabla_{\tilde{\bm\eta}^{t}}\rho_{t}^{2}\right\Vert_{2}^{2}\right] - \mathbb{E}[r_{t+1}].\label{eq:eq-11}
\end{align}
Therefore, for any $T\geq 1$, we always have
\begin{align}
 &\mathbb{E}\left[\gamma^{T}\rho^{2}(\bm q_{T+2},\tilde{\bm\eta}^{T+1}) - \rho^{2}(\bm q_{2},\tilde{\bm\eta}^{1})\right]\notag \\
 &\quad \leq - \sum_{t=1}^{T}\gamma^{t}\left(\tilde{\alpha}^{t}
 \mathbb{E}\left[\left\Vert \nabla_{\tilde{\bm\eta}^{t}}\rho_{t}^{2}\right\Vert_{2}^{2}\right] + \mathbb{E}[r_{t+1}]\right).\label{eq:eq-12}
\end{align}

Re-arranging the terms in (\ref{eq:eq-12}), it follows that
\begin{align}
 & \sum_{t=0}^{T}\gamma^{t}\tilde{\alpha}^{t}\mathbb{E}
 \left[\left\Vert\nabla_{\tilde{\bm\eta}^{t}}\rho_{t}^{2}\right\Vert_{2}^{2}\right]
 \leq\mathbb{E}\left[\rho^{2}(\bm q_{2},\tilde{\bm\eta}^{1})\right].
\end{align}

Similarly, let the GD step-size $\alpha^{t}$ be chosen such that $\tilde{\alpha}^{t}=\alpha^{t}-\frac{1}{2}\tilde{L}(\alpha^{t})^{2}=\frac{1}{t}$. By taking the limit $T\rightarrow\infty$, it follows that
\begin{align}\label{eq:eq-14}
 \lim_{T\rightarrow\infty}\Big(\sum_{t=1}^{T}\frac{\gamma^{t}}{t}\Big)
 \liminf_{t\geq0}\mathbb{E}\left[\left\Vert\nabla_{\tilde{\bm\eta}^{t}}\rho_{t}^{2}\right\Vert_{2}^{2}\right]
 \leq \mathbb{E}\left[\rho^{2}(\bm q_{2},\tilde{\bm\eta}^{1})\right].
\end{align}

Therefore, it follows that
\begin{align}\label{eq:eq-15}
 & \liminf_{t\geq0}\mathbb{E}\left[\left\Vert \nabla_{\tilde{\bm\eta}^{t}}\rho_{t}^{2}\right\Vert_{2}^{2}\right]
 \leq\frac{\mathbb{E}\left[\rho^{2}(\bm q_{2},\tilde{\bm\eta}^{1})\right]}
 {\mathrm{E_{1}}\left(\ln\frac{1}{\gamma}\right)},
\end{align}
which completes the proof of the convergence of the GD method at the MEC server.


\end{document}